\documentclass[preprint,a4paper,onecolumn,
notitlepage,secnumarabic,amssymb,aps,prb,superscriptaddress,floatfix]{revtex4-2}

\usepackage{natbib}
\usepackage{balance}
\usepackage[T1]{fontenc}
\usepackage[compact]{titlesec}
\usepackage{graphicx}
\usepackage{svg}
\usepackage{amsmath}
\usepackage{subfigure}
\usepackage{braket}
\usepackage{hyperref}
\usepackage{bm}
\hypersetup{
    colorlinks=true,
    linkcolor=blue,
    filecolor=red,      
    urlcolor=blue,
    citecolor=magenta
}
\usepackage[version=4]{mhchem}
\usepackage{ragged2e}
\usepackage{float}

\begin{document}

\title{Hybrid Quantum--Classical Machine Learning Potential with Variational Quantum Circuits}

\author{Soohaeng Yoo Willow}
\affiliation{Department of Energy Science, Sungkyunkwan University, Seobu-ro 2066, Suwon, 16419, Korea}

\author{David ChangMo Yang}
\affiliation{Department of Energy Science, Sungkyunkwan University, Seobu-ro 2066, Suwon, 16419, Korea}

\author{Chang Woo Myung}
\email{cwmyung@skku.edu}
\affiliation{Department of Energy Science, Sungkyunkwan University, Seobu-ro 2066, Suwon, 16419, Korea}
\affiliation{Department of Energy, Sungkyunkwan University, Seobu-ro 2066, Suwon, 16419, Korea}
\affiliation{Department of Quantum Information Engineering, Sungkyunkwan University, Seobu-ro 2066, Suwon, 16419, Korea}

\date{August 6, 2025}

\begin{abstract}
Quantum algorithms for simulating large and complex molecular systems are still in their infancy, and surpassing state-of-the-art classical techniques remains an ever-receding goal post. A promising avenue of inquiry in the meanwhile is to seek practical advantages through hybrid quantum-classical algorithms, which combine conventional neural networks with variational quantum circuits (VQCs) running on today's noisy intermediate-scale quantum~(NISQ) hardware.
Such hybrids are well suited to NISQ hardware. The classical processor performs the bulk of the computation, while the quantum processor executes targeted sub-tasks that supply additional non-linearity and expressivity. Here, we benchmark a purely classical $E(3)$-equivariant message-passing machine learning potential (MLP) against a hybrid quantum-classical MLP for predicting density functional theory (DFT) properties of liquid silicon. In our hybrid architecture, every readout in the message-passing layers is replaced by a VQC.
Molecular dynamics simulations driven by the HQC-MLP reveal that an accurate reproduction of high-temperature structural and thermodynamic properties is achieved with VQCs. These findings demonstrate a concrete scenario in which NISQ-compatible HQC algorithm could deliver a measurable benefit over the best available classical alternative, suggesting a viable pathway toward near-term quantum advantage in materials modeling.
\end{abstract}

\maketitle

\newpage
\section{Introduction}
Quantum Machine Learning~(QML) has emerged as a significant research frontier, aiming to integrate the principles of quantum computation with the adaptive frameworks of machine learning. The central premise of QML is its potential to leverage quantum phenomena, such as superposition and entanglement, to process information in high-dimensional Hilbert spaces. This capability is theorized to offer substantial advantages over classical models for tasks involving intricate data correlations, leading to active investigation into its application for optimization, classification, sampling techniques, and molecular modeling~\cite{biamonteQuantumMachineLearning2017,sajjanQuantumMachineLearning2022,zamanSurveyQuantumMachine2024,zaman_comparative_2025,tillyVariationalQuantumEigensolver2022,huangTermQuantumComputing2023,mariTransferLearningHybrid2020,bowlesBetterClassicalSubtle2024}.

However, despite the growing enthusiasm surrounding QML, there is also skepticism about its practical utility, especially in the near term. Many quantum models have struggled to outperform their classical counterparts in key areas, and the exact role of ``quantumness'' in improving performance remains an open question. Studies have shown that while quantum circuits can represent certain nonlinear relationships more efficiently, the inductive bias provided by quantum systems and the problems for which they are particularly well suited are still not fully understood~\cite{West2024}. Classical baseline models outperform prototypical quantum models in small-scale benchmark datasets and no clear advantage of quantum entanglement in QML was found~\cite{West2024}. As such, QML research remains in a stage of exploration, with both its potential and limitations yet to be fully defined. 

One of the most compelling and underexplored of these domains is the development of machine learning potentials (MLPs) for materials science~\cite{thiemann_introduction_2025_review,aldossary_silico_2024_review,wan_construction_2024_review,kocer_neural_2022_review,unke_mlff_2021_review,behler_four_2021_review,dral_quantum_2020_review,merchant_scaling_2023,batatia_foundation_2024,myung_challenges_2022}.
This field is particularly suitable for quantum approaches because the underlying atoms and molecules being modeled are inherently quantum mechanical. The main challenge in this area has always been the rather unfavorable computational cost of traditional \textit{ab initio} methods like density functional theory (DFT), Møller-Plesset~(MP) perturbation theory, and coupled cluster~(CC) theory.  In particular, the conventional DFT method scales with the system size $N$ as $\mathcal{O}(N^3)$, while second-order MP theory~(MP2) and CC theory with single, double, and perturbative triple excitations~[CCSD(T)] scale steeply as $\mathcal{O}(N^5)$ and $\mathcal{O}(N^7)$, respectively. As capable as the latter methods are of capturing complex quantum mechanical correlations for accurate description of molecular systems, their scaling laws render large-scale simulations intractable. To address this scaling problem, classical MLPs have emerged as a revolutionary tool over the past two decades, capable of learning the complex potential energy surface from reference \textit{ab initio} calculations, enabling simulations with near-quantum accuracy at a fraction of the computational cost. The field has seen rapid progress, evolving from architectures like feed-forward neural networks~\cite{behlerGeneralizedNeuralNetworkRepresentation2007,eckhoffHighdimensionalNeuralNetwork2021} and Gaussian Approximation Potentials~\cite{bartok_gap_2010,bartok_representing_2013,bartok_gaussian_2015,klawohn_gap_2023,hajibabaei_machine_2021,hajibabaei_sparse_2021,hajibabaei_universal_2021,ha_sparse_2022,willow_active_2024,quinonero-candela_unifying_2005} to the current state-of-the-art $E(3)$-equivariant graph neural networks, which respect the fundamental symmetries of physics~\cite{e3nn_paper,batzner_e3-equivariant_2022,cohen_general_2019,cohen_gauge_2019}. Despite their remarkable success in enabling large-scale materials simulations, these classical models are not without their own limitations. Their accuracy is highly dependent on extensive, computationally expensive training datasets, and they still face challenges in capturing the non-local correlations present in complex systems. This creates a clear and compelling opportunity to investigate whether the integration of QML can push the boundaries of what is possible in atomistic modeling.

The potential for quantum advantage in QML is often linked to the ability to recognize hidden patterns in quantum data. Non-local quantum measurements can extract underlying parameters from quantum datasets using only a few samples~\cite{cerezo_challenges_2022}. In the context of applying QML to MLPs, referred to as quantum MLPs (QMLPs), the primary goal is to demonstrate quantum advantage by surpassing the performance of classical state-of-the-art MLP models. However, the challenge arises from the fact that the local chemical environment, which is inherently classical, must be mapped onto qubits. This poses a disadvantage, as no known quantum speedup exists for such classical-to-quantum data transformations~\cite{shiota_universal_2024,couzinie_towards_2025,le_gall_robust_2025}. Despite this challenge, several studies suggest that polynomial advantage may still be possible for purely classical problems using quantum algorithms~\cite{cerezo_challenges_2022,chia_sampling-based_2020,babbush_focus_2021,grover_fast_1996,Bernstein_1997}. However, a clear understanding of how quantum advantage would manifest itself in classical tasks such as QMLP has yet to emerge. For this reason, it is crucial to investigate potential quantum advantages in two key areas: data embedding and the variational Ansatz. When classical local chemical environments are embedded into qubits, the effectiveness of the embedding scheme becomes critical. Furthermore, understanding whether exploiting quantum entanglement can improve machine learning performance during training is another essential aspect. The construction of an effective variational Ansatz circuit with activation functions for potential energy and force regression is crucial for exploring the quantum advantage in QMLPs~\cite{zi_efficient_2024}. Building an equivariant QNN is also important for probing the quantum advantage in QMLPs~\cite{West2024,couzinie_towards_2025}. These aspects require careful consideration to elucidate how quantum resources may offer tangible benefits over classical approaches.

In contrast to the considerable attention QML has received in other fields, its application to MLPs in materials science remains largely unexplored. Given the quantum nature of the systems being modeled, there is an untapped potential for QMLPs to enhance the efficiency and accuracy of MLPs. The hybrid quantum-classical MLP~(HQC-MLP), which combines the strengths of quantum and classical computing, is particularly well suited for this task. HQC-MLP leverages the processing power of quantum circuits while utilizing the stability and scalability of classical models, offering a balanced approach that can take advantage of current quantum hardware without full quantum encoding of material descriptors~\cite{biamonteQuantumMachineLearning2017,sajjanQuantumMachineLearning2022}.

Here, we develop HQC-MLP methods and apply them for the simulation of molecular dynamics (MD) of liquid silicon at a computationally frequented~\cite{stillinger_computer_1985,Lenosky_2000,zhang_polymorphism_2015} temperature of 2000~K and an unreported temperature of 3000~K. Our HQC-MLP models are trained with the results from \textit{ab initio} MD~(AIMD) simulation of liquid silicon.
The classical processor performs the bulk of the computation, while the quantum processor executes targeted sub-tasks that supply additional non-linearity and expressivity.
We find that this hybrid approach not only accelerates the training process, but also achieves predictive accuracy comparable to purely classical models, suggesting that HQC-MLP holds significant potential for advancing the field of materials science through the efficient modeling of complex atomistic systems.

\begin{figure}[htp]
\centering
\includegraphics[width=\linewidth]{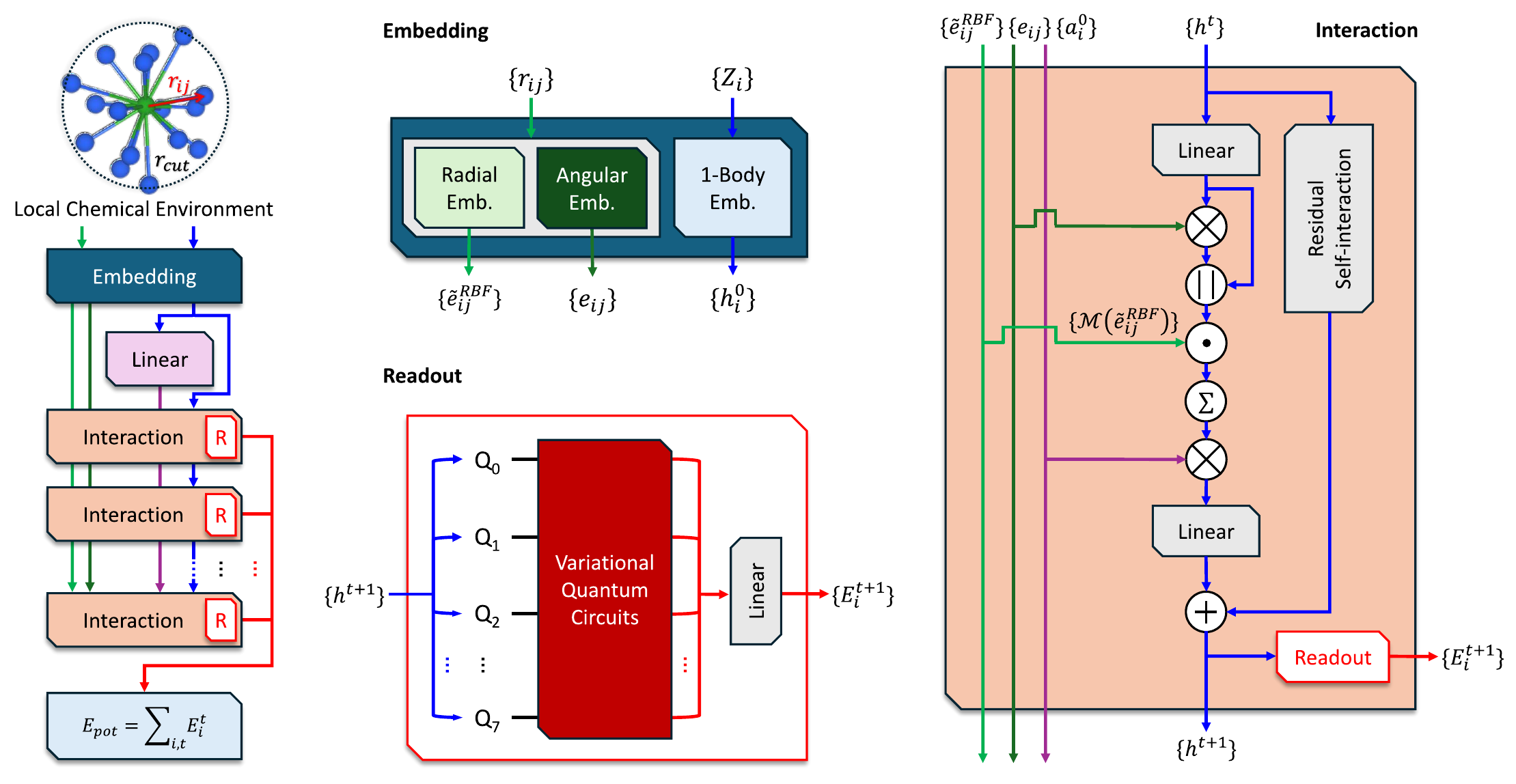}
\caption{Architecture of the hybrid quantum-classical machine learning potential~(HQC-MLP). The HQC-MLP model consists of an embedding layer and interaction layers. Each interaction layer includes a readout layer that predicts output values. In the embedding layer, node features ($h_i$) are initialized based on the atomic numbers of the nodes $\{Z_i\}$, while edge features $e_{ij}$ and radial basis vectors $\tilde{e}_{ij}^\text{RBF}$ are assigned based on pairwise vectors $\bm{r}_{ij} = \bm{r}_j - \bm{r}_i$. The interaction layer updates node features in a \textsc{ResNet}-like scheme. The readout layer applies a nonlinear transformation to the node features, generating output values that represent the energy value associated with each node. In the HQC-MLP model, variational quantum circuits \texttt{VQC}, executing nonlinear transformations~\cite{sajjanQuantumMachineLearning2022,zamanSurveyQuantumMachine2024}. It includes a feature map module, which takes the node features as input, and an Ansatz module, which contains trainable parameters~${\theta}$.  Five different \texttt{VQC} Ansatz modules in Figure~\ref{fig:Ansatz} were tested.
}
\label{fig:hybrid_model}
\end{figure}

\section{Hybrid quantum-classical machine learning potential}

The objective of our model is to learn a mapping from the positions $\{\bm{r}_i\}$ and chemical species $\{Z_i\}$ of the $i$-th atom of a system to its total potential energy
\begin{align}
    E_\text{pot} & = \sum_{i}^{N_\text{atoms}}\sum_{t}^{N_\text{layers}} E_i^{(t)} = \sum_i^{N_\text{atoms}} E_i
\end{align}
and the corresponding atomic forces $\bm{f}_i = -\nabla_i E_\text{pot}$. The model is designed to be energy-conserving by the definition of forces as the negative gradient of the potential energy. 

The total potential energy is decomposed into the sum of local atomic contributions $E_i = \sum_t ^{N_\text{layers}} E_i ^{(t)}$. Crucially, the model must adhere to physical symmetries: $E_\text{pot}$ must be invariant under translations, rotations, and reflections of the system, while vector quantities like forces and internal geometric features must be correspondingly equivariant.

The architecture of our HQC-MLP (Figure~\ref{fig:hybrid_model}) is based on an $E(3)$-equivariant message passing neural network (MPNN). In this framework, each atomic structure is converted into a graph where atoms are nodes and the edges connect neighbors within a cutoff radius~$r_\text{cut}$. MPNN iteratively updates the node features through message-passing operations. To ensure the required symmetries, the message-passing convolutions employ steerable filters, $S(\bm{r}_{ij})$, constructed from products of learnable radial functions~$R(r_{ij})$ and spherical harmonics~$Y_m ^{(l)} (\hat{\bm{r}}_{ij})$:
\begin{align}
    \label{eq:S_ml}
    S_m^{(l)} (\bm{r}_{ij}) & = R^{(l)}(r_{ij}) Y_m^{(l)} (\hat{\bm{r}}_{ij})
\end{align}
where $\bm{r}_{ij} = \bm{r}_j - \bm{r}_i$ are the pairwise relative interatomic coordinates which can be represented by distances~$r_{ij}$ and unit directional vectors~$\hat{\bm{r}}_{ij}$. The irrep index $(l)$, also known as the angular momentum quantum number, is understood as the layer index. This construction makes the filters, and consequently the network features, equivariant under $SO(3)$ rotations.

The features of the atoms are represented as learnable node features ($h_i^{(t)} \in \mathbb{R}^F$), where $F$ represents the dimension of the node features
and $(t)$ is the index of the (composite) interaction layer. 
The features of the initial node $h_i^{(0)}$ are determined based on an embedding related to the atomic numbers $\{Z_i\}$ as $h_i^{(0)} = \bm{a}_{Z_i}$, where the atomic-type embeddings $\bm{a}_{Z}$ are randomly initialized and then optimized during the training process.

The radial basis vectors $\tilde{e}_{ij}^\text{RBF} \in \mathbb{R}^{N_\text{RBF}}$ for the interatomic distance~\cite{gasteiger_directional_2021} are defined by Bessel basis functions $B_n (x) = \sqrt{\frac{2}{r_c^3}} \frac{j_0 ( n \pi x))}{|j_1 (n \pi)|}$ and a polynomial envelope function $f_\text{env}$ as
\begin{align} \label{eq:RBF}
    \tilde{e}_\text{RBF} (r_{ij}) & = B_n (r_{ij} / r_c) f_\text{env} (r_{ij}, r_c) \; .
\end{align}
Here, $j_0$ and $j_1$ are spherical Bessel functions of the first kind. The edge features $e_{ij}$ encode the direction information and are obtained as $Y_m^{(l)} (\hat{r}_{ij})$.

The goal of the interaction layer is to update the node energy value by incorporating the effects of edge interactions, including the bond and dispersion energies. 
To achieve this, the node features are updated in the scheme of \textsc{ResNet}~\cite{he_deep_2016}:
\begin{align}
\bm{h}^{(t+1)} & = f_\text{SI} (\bm{h}^{(t)}) + f (\bm{h}^{(t)}, \bm{e}_{ij}, \tilde{\bm{e}}_{ij}^\text{RBF}) \; . 
\end{align}
Here, $f$ represents a sequence of an atom-wise dense layer $f_\text{dense}$, an interatomic continuous-filter convolution layer $f_\text{conv}$, and two additional atom-wise dense layers. 
The weights in the self-interaction layer $f_\text{SI}$ are learned separately for each atomic number.

In the convolution layer $f_\text{conv}$, the node features are updated according to
\begin{align} \label{eq:conv}
    h_i & = 
    \sum_{j \in N_i} \left[ (e_{ij} \otimes h_j) \odot \mathcal{M}(\tilde{e}_{ij}^\text{RBF}) \right]
\end{align}
where $\otimes$ denotes the rotationally equivariant tensor product (as detailed in the Supporting Information, Eq.~S1)
and $\odot$ refers to the element-wise multiplication. 
A multilayer perceptron $\mathcal{M}$ is applied to the radial basis vectors $\tilde{e}_{ij}^\text{RBF}$ to capture interatomic interactions based on distance. 
As a result, $\mathcal{M}(\tilde{e}_{ij}^\text{RBF})$ encodes the learnable radial functions $R^{(t)}(r_{ij})$ in Eq.~\ref{eq:S_ml}. 
Finally, the convolution filters $S_m^{(t)}$ in Eq.~\ref{eq:S_ml} are encoded through the convolution layer $f_\text{conv}$ within the models.

\begin{figure} [htp]
\centering
\includegraphics[width=\linewidth]{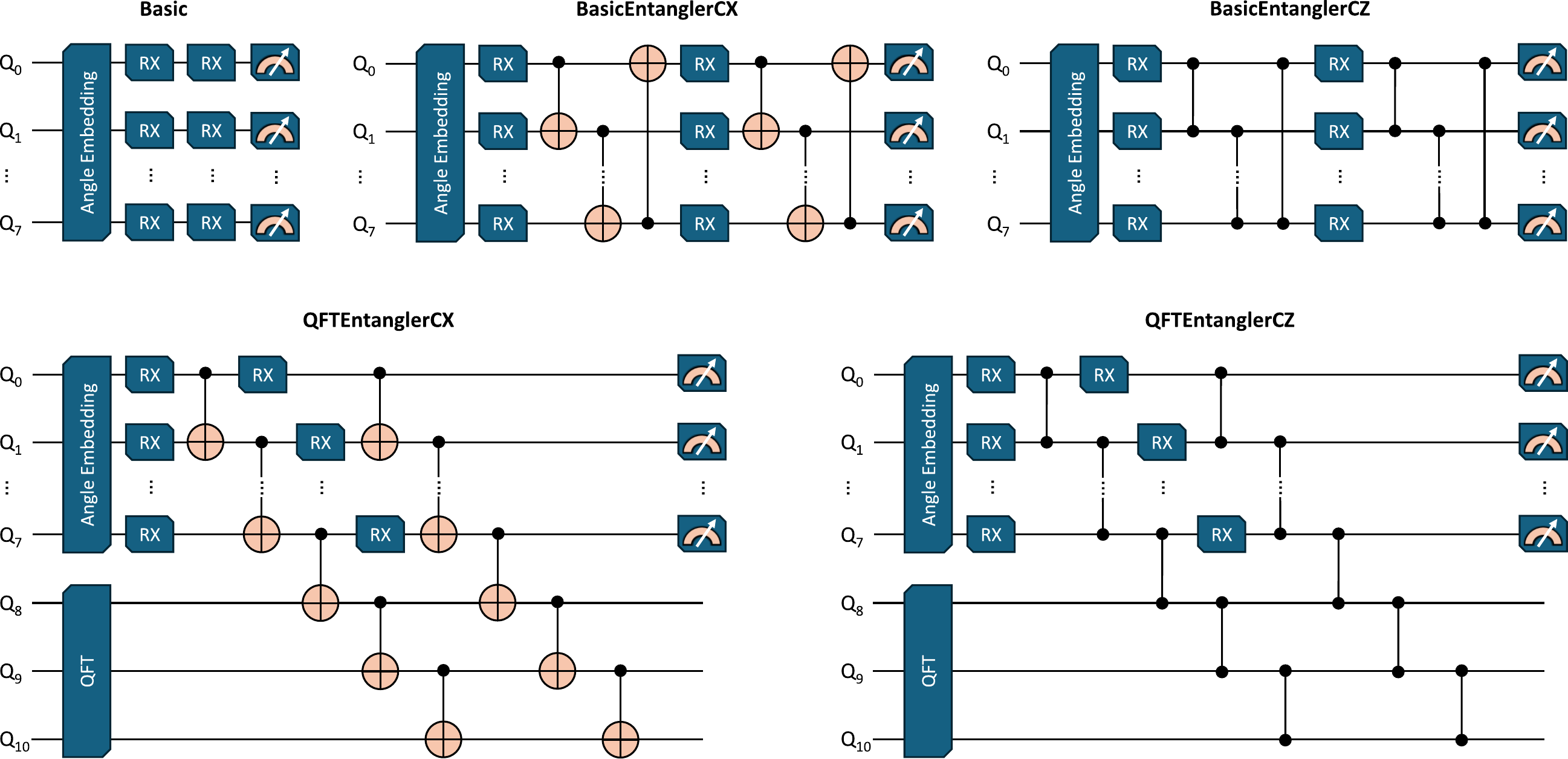}
\caption{Diagrams of five distinct Ansatz modules with 8 qubits for \texttt{AngleEmbedding}, 3 qubits for \texttt{QFT}, and 2 layers for trainable rotational gates \texttt{RX}.  
The \texttt{AngleEmbedding} module served as the feature map across all quantum circuits, each using one of five Ansatz module variations. 
These Ansatz modules are categorized based on the type of the entanglement and whether quantum Fourier transform (\texttt{QFT}) is incorporated.
Entanglement between qubits is achieved using either the controlled-NOT~(\texttt{CX}) or the controlled-$Z$~(\texttt{CZ}). All Ansatz modules have three layers.
\label{fig:Ansatz}
}

\end{figure}

To calculate the energy value $E_i^{(t)}$ of node $i$ from its feature at the $t$-th interaction layer, the invariant part of the node feature is mapped to the node energy via the readout function:
\begin{align}
    E_i^{(t)} & = \sum_{\tilde{k}} W^{(t)}_{\tilde{k}}~ (h_{i,\tilde{k}}^{(t)}),
\end{align}
where $W_{\tilde{k}}^{(t)}$ represents the readout weights, and $h_{i, \tilde{k}}^{(t)}$ represents the $\tilde{k}$-th element of the $i$-th node feature at the current layer $(t)$.
To ensure invariance of the node energy $E_i^{(t)}$,
the readout layer is only based on the invariant features (those with rotational order of $l=0$).
These features are obtained by applying the transformation $h_i^{(t)} = \text{Gate}(f_\text{dense}^{l=0} (h_i^{(t)})$), where $\text{Gate}(\cdot)$ refers to an equivariant \texttt{SiLU}-based gate in the MLP on classical computers. In our HQC-MLP model, we replace this classical readout part with variational quantum circuits~(VQCs). The invariant feature vector of the classical neural network is used as input to parameterize the VQCs. The circuit's output is interpreted as the local energy $E_i^{(t)}$. This substitution allows a quantum processor to handle the final nonlinear mapping from the learned latent features to local energy, representing the core of our hybrid approach.

The \texttt{VQC} module is designed to map the classical input features to the final energy contribution and consists of three stages: feature encoding, a variational Ansatz, and measurement. First, the invariant node features from the message-passing, $h_{i,\tilde{k}}^{(t)}$, are encoded onto an $N$-qubit state using an angle embedding scheme. Each element of the feature vector is used to parameterize a single-qubit rotation gate (e.g.~\texttt{RY} gates), which prepares the initial state for the variational circuit. The core of the \texttt{VQC} is the parameterized Ansatz, which performs a nonlinear transformation on the encoded state. To systematically study the impact of circuit structure on model performance, we designed and tested five Ansatz architectures~(Figure~\ref{fig:Ansatz}). These architectures are differentiated by their primary components, specifically the type of entangling gates used (\texttt{CX}~vs.~\texttt{CZ}) and the optional inclusion of a quantum Fourier transform~(\texttt{QFT}) block. For instance, we designed a baseline circuit with a simplified structure in order to isolate the effect of entanglement, which is a key aspect of ``quantumness''. Each Ansatz was constructed with three repeated layers, a depth found to be effective in related QML applications~\cite{zaman_comparative_2025}. The trainable parameters of the Ansatz are optimized concurrently with the classical components of the model. Finally, the energy contribution is extracted by measuring the expectation value of the Pauli~$Z$ observable. All quantum circuits were implemented using the \textsc{PennyLane} software library~\cite{bergholmPennyLaneAutomaticDifferentiation2022}. Our simulations were conducted using a total of 8 qubits for the main circuit, with the \texttt{QFT} module, when present, acting on an additional 3-qubit subsystem.

\section{Results and discussion}

\begin{figure}[htp]
\centering
 \includegraphics[width=0.9\linewidth]{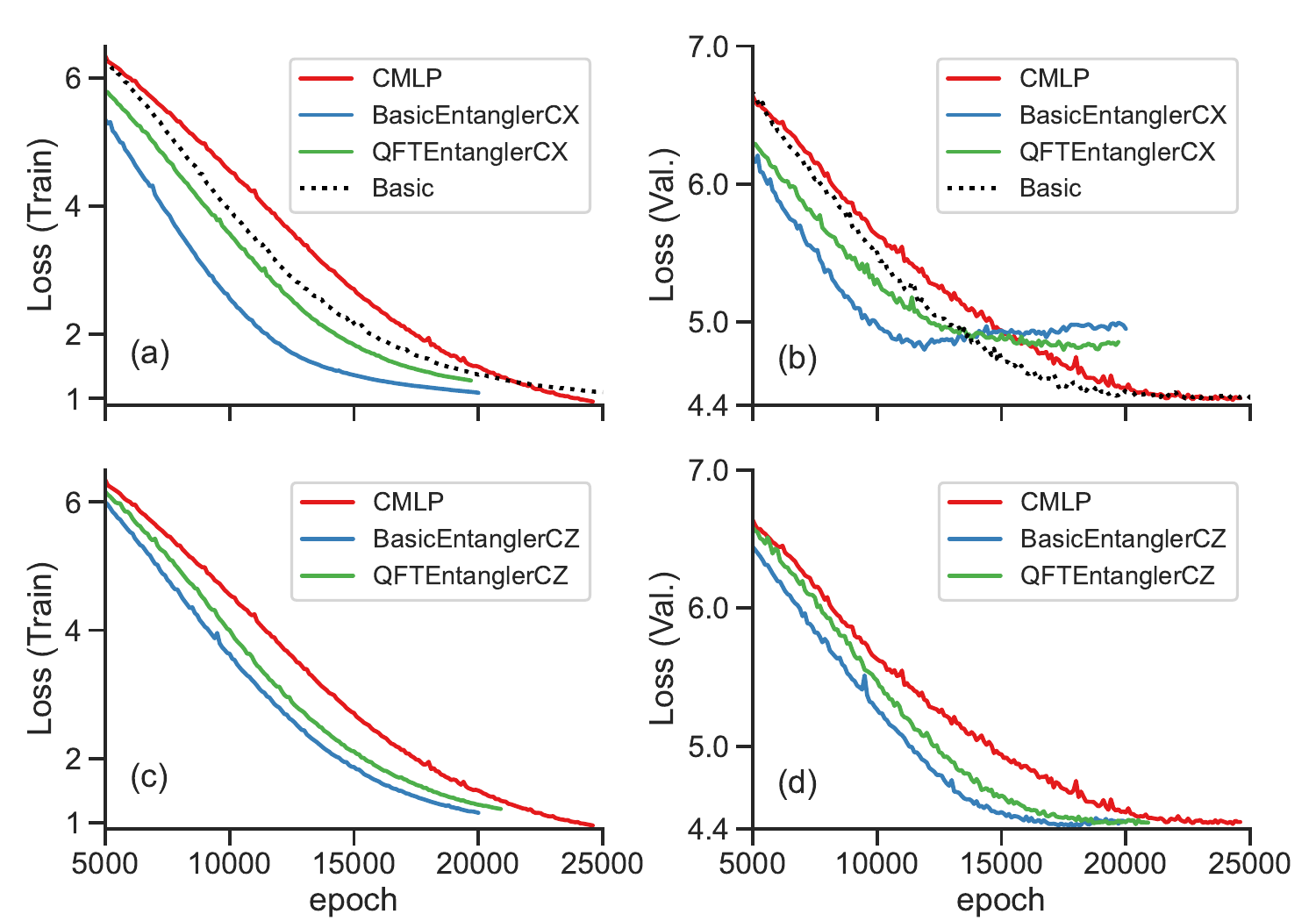}
\caption{Loss profiles for hybrid quantum-classical machine learning potential~(HQC-MLP) models in comparison to a classical machine learning potential~(CMLP) model. 
Loss values are evaluated for (a, c) training and (b, d) validation datasets.
In the training dataset, loss values continue to decrease.
In the validation dateset, however, loss values reach minimum points, indicating the onset of model overfitting to the training data. }
\label{fig:loss}
\end{figure}

\subsection{Training of HQC-MLP models}

To evaluate the models, we generated a dataset of 1,000 liquid silicon configurations at 3000~K from Born--Oppenheimer molecular dynamics~(BOMD) simulations using DFT. This was partitioned into 900+50+50 configurations for training, validation, and test, respectively. Figure~\ref{fig:loss} compares the training and validation loss curves for the CMLP and the HQC-MLP models.

A significant challenge for scaling \texttt{VQCs} is the issue of barren plateaus (BPs) --- the exponential vanishing of gradients during training~\cite{mcclean_barren_2018}. This phenomenon is attributed to several factors, including high circuit expressivity, the use of global loss functions, and, notably, the volume of entanglement in the circuit~\cite{Holmes2022Connecting, cerezo_cost_2021, Sharma2022Trainability, patti_entanglement_2021}.
Although BPs typically manifest as failure to train, the associated high expressivity and entanglement can also lead to poor generalization.
Figures \ref{fig:loss} and S1 show that, compared to the CMLP model, HQC-MLP models which incorporate entanglement exhibit a rapid decrease in training loss followed by an early onset of overfitting.

A primary observation is that the HQC-MLP models consistently converge faster on the training data than the purely classical model. This acceleration is particularly pronounced for models that incorporate entangling gates, such as the controlled-NOT~(\texttt{CX}). However, these hybrid models also exhibit an earlier onset of overfitting, indicated by a premature increase in their validation loss. This tendency was evident in the models utilizing \texttt{CX} entanglement. In contrast, the inclusion of a \texttt{QFT} block for 2D rotation equivariant circuit~\cite{West2024} appears to improve generalization, resulting in lower final loss values on the validation set. These findings suggest that while entanglement --- a key aspect of ``quantumness'' --- can accelerate fitting to the training data, other architectural equivariant components like the \texttt{QFT} may be crucial for enhancing the model's predictive accuracy on unseen data.

Our results reveal a complex interplay between architectural choice and model performance. A consistent advantage across all HQC-MLP models was an accelerated training convergence compared to the state-of-the-art CMLP. However, this speed was often accompanied by a tendency to overfit, motivating a search for the \texttt{VQC} architecture with the best generalization performance. Our investigation into different \texttt{VQC} designs identified the choice of entangling gate as the most critical factor. The models utilizing the controlled-$Z$~(\texttt{CZ}) gate consistently achieved the best performance, yielding the lowest validation loss among all tested architectures. This superior result was robust, holding for both the simple \texttt{BasicEntanglerCZ} model and the partially 2D equivariant \texttt{QFTEntanglerCZ} model~\cite{West2024}. In contrast, the models based on the \texttt{CX} gate, while also training rapidly, struggled with more severe overfitting. The addition of the \textsc{QFT} block prevented overfitting but did not fully remedy this issue for the \texttt{CX} variants, whose performance remained inferior to the \texttt{CZ}-based circuits. This distinct performance gap suggests that the nature of the entanglement operation is paramount for this task. The \texttt{CZ} gate, which imparts a conditional phase, appears to be a more effective strategy for this regression problem than the state-flipping \texttt{CX} gate. Therefore, while a definitive ``quantum advantage'' remains an open question, our findings point to two clear conclusions: a practical benefit in training speed for all hybrid models, and a distinct performance advantage achieved by using \texttt{CZ}-based entanglers. This suggests that for this problem, the introduction of entanglement via these circuit designs may hinder generalization more than the partial restoration of symmetry helps. Consequently, future work should explore more structured entanglement strategies, which have been theorized to mitigate BPs~\cite{patti_entanglement_2021} and could potentially lead to models that are both expressive and generalizable.

The root mean squared errors (RMSE) for the predicted energies (in meV/atom) and forces (in eV/\AA) from the models were estimated using the test dataset, as shown in Figure~S2.
Both the HQC-MLP and CMLP models produced an RMSE of approximately 2.0~meV per atom for the predicted energies ($\text{RMSE}_E$). 
Similarly, the RMSE for the predicted forces ($\text{RMSE}_F$) was about 0.2~eV/\AA. 
These RMSE values are comparable to the results reported by Behler and Parrinello~\cite{behlerGeneralizedNeuralNetworkRepresentation2007}, where $\text{RMSE}_E$ was around 4~meV per atom, and $\text{RMSE}_F$ was approximately 0.2~eV/\AA.

\subsection{Molecular dynamics simulations of liquid silicon with HQC-MLP models}

Simulating liquid silicon at high temperatures is a well-known challenge, as standard empirical potentials like those of Bazant~\cite{Bazant_Si}, Lenosky~\cite{Lenosky_2000}, and Tersoff~\cite{Tersoff_Si} are known to inadequately describe its physical properties. DFT, in contrast, provides an accurate reference that agrees well with experimental data~\cite{behlerGeneralizedNeuralNetworkRepresentation2007}. To assess the ability of our HQC-MLP and compare against CMLP models in this demanding regime, we performed 100~ps molecular dynamics~(MD) simulations of liquid silicon at $T = 2000~\text{K}$. A crucial first test for any new potential is the conservation of total energy. We confirmed that all tested HQC-MLP and CMLP models were stable, conserving the total energy throughout the full simulation trajectory, which indicates a smooth learned potential energy surface and accurate forces (see Figure~S3).

\begin{figure}[htp]
\centering
\includegraphics[width=10cm]{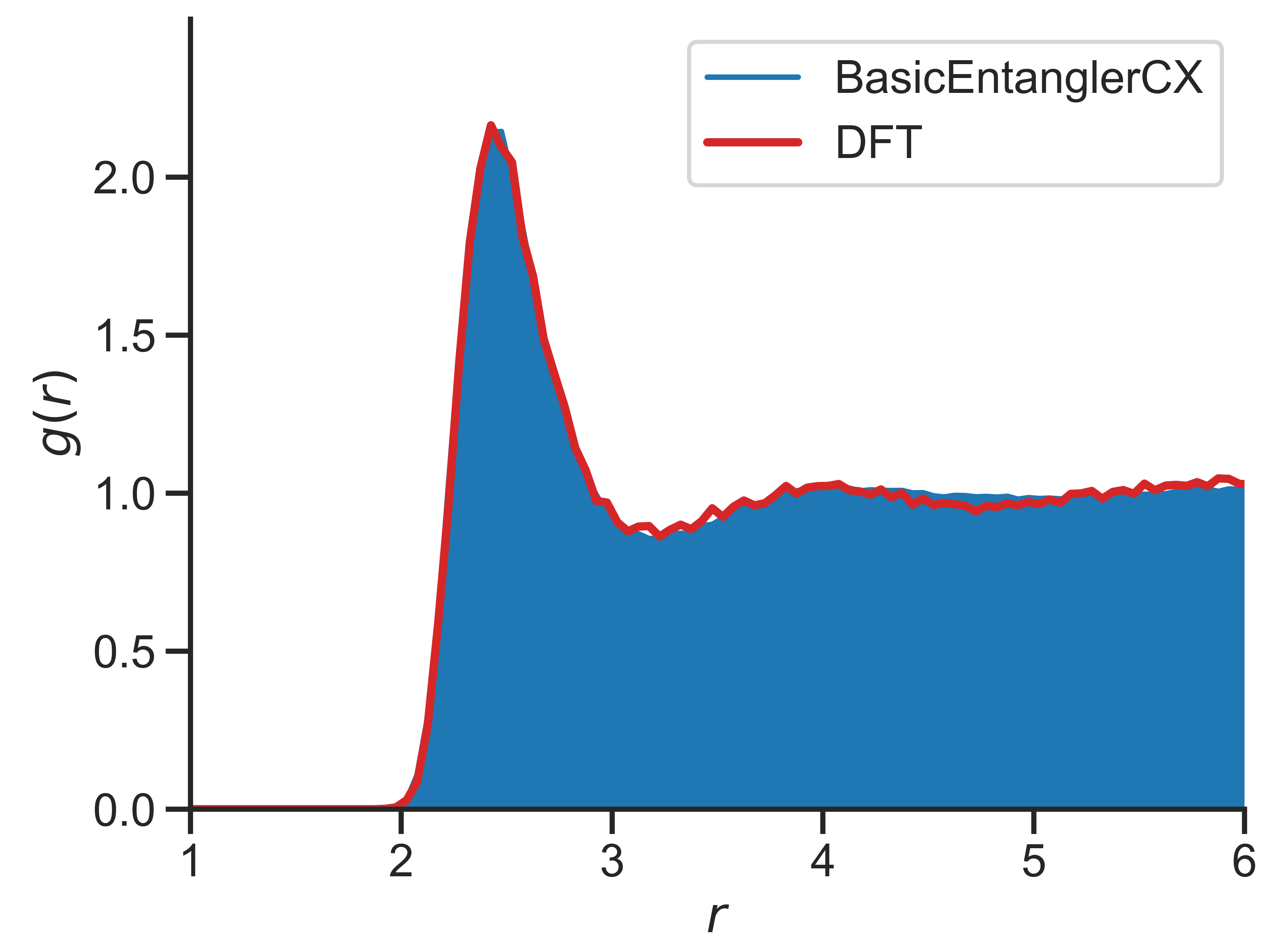}
\caption{Radial distribution function, $g(r)$, for liquid silicon at $T = 2000~\text{K}$. The plot compares the result from the HQC-MLP(\texttt{BasicEntanglerCX}) model (blue filled area) with the reference calculated using DFT (red line), demonstrating excellent agreement. Both curves were obtained from MD simulations of a 64-atom system in a cubic cell ($a = 10.862$~\AA).}
\label{fig:RDF}
\end{figure}

Figure~\ref{fig:RDF} shows the radial distribution function (RDF) for liquid silicon at $T = 2000 ~\text{K}$, comparing results from the HQC-MLP(\texttt{BasicEntanglerCX}) model with those from the DFT simulations.
The DFT-based BOMD simulations were performed for 10 ps to reach equilibrium, followed by 18 ps for data production.
In contrast, MD simulations using the HQC-MLP(\texttt{BasicEntanglerCX}) model were run for 100 ps. 
The RDF obtained with the HQC-MLP(\texttt{BasicEntanglerCX}) model closely aligned with the RDF produced by the DFT calculations.

\begin{figure}[htp]
\centering
\includegraphics[width=12cm]{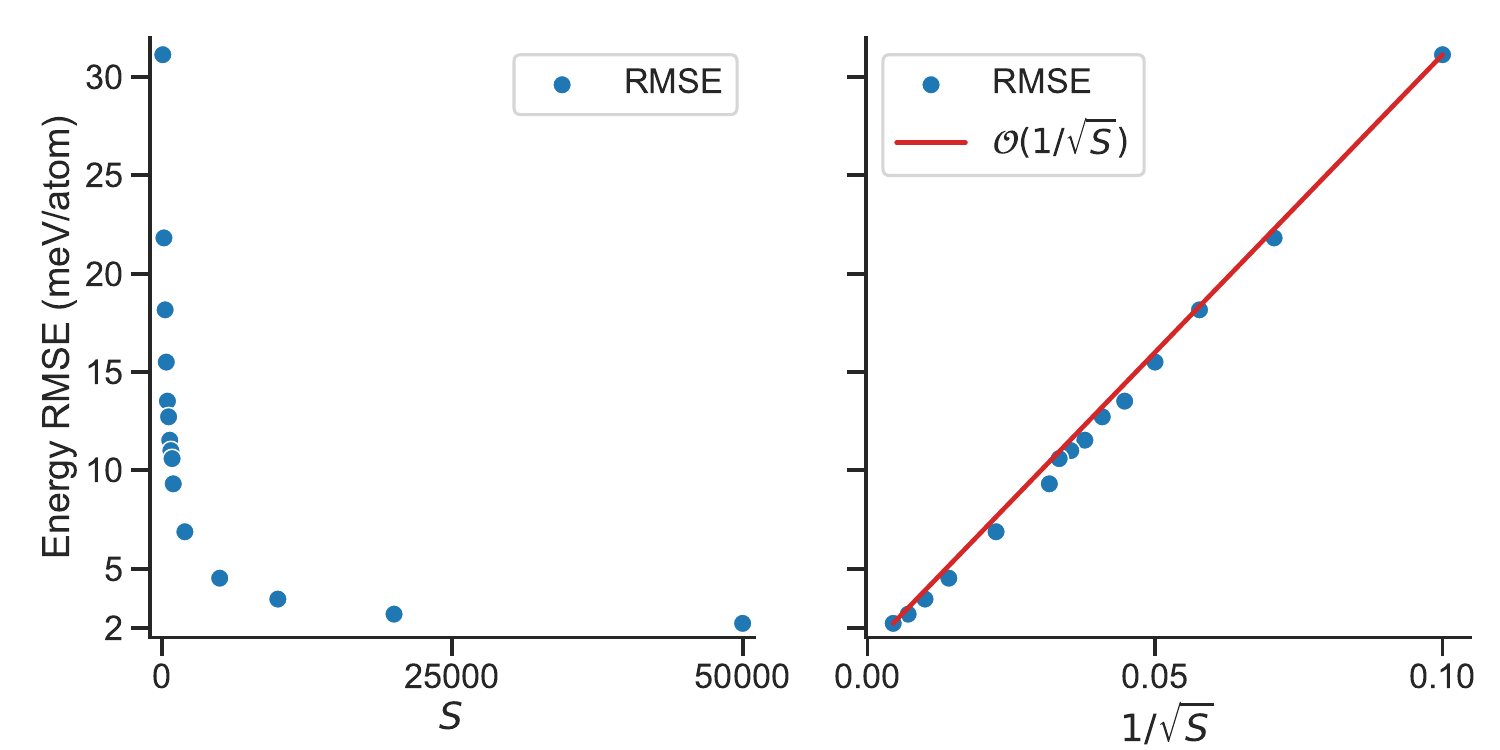}
\caption{Dependence of HQC-MLP predictions on the number of measurement shots $S$. Root-mean-square-error (RMSE) decreases with $\mathcal{O} (1/\sqrt{S}$).}
\label{fig:shots}
\end{figure}
\subsection{Effect of Finite Measurement Shots}

The preceding analyses were based on mathematically exact expectation values from the VQCs. To assess the model's performance in realistic quantum hardware, we investigate the impact of statistical noise arising from a finite number of measurement shots, $S$. Figure~\ref{fig:shots} quantifies the convergence of the predicted potential energies as a function of $1/\sqrt {S}$. As expected from sampling theory, the root mean square error (RMSE) between the shot-based predictions and the exact expectation values scales as $\mathcal{O}(1/\sqrt{S})$. We determined that achieving an RMSE of approximately 2.0~meV/atom requires on the order of $5 \times 10^4$ measurement shots per energy evaluation.

\begin{figure}[htp]
\centering
\includegraphics[width=8cm]{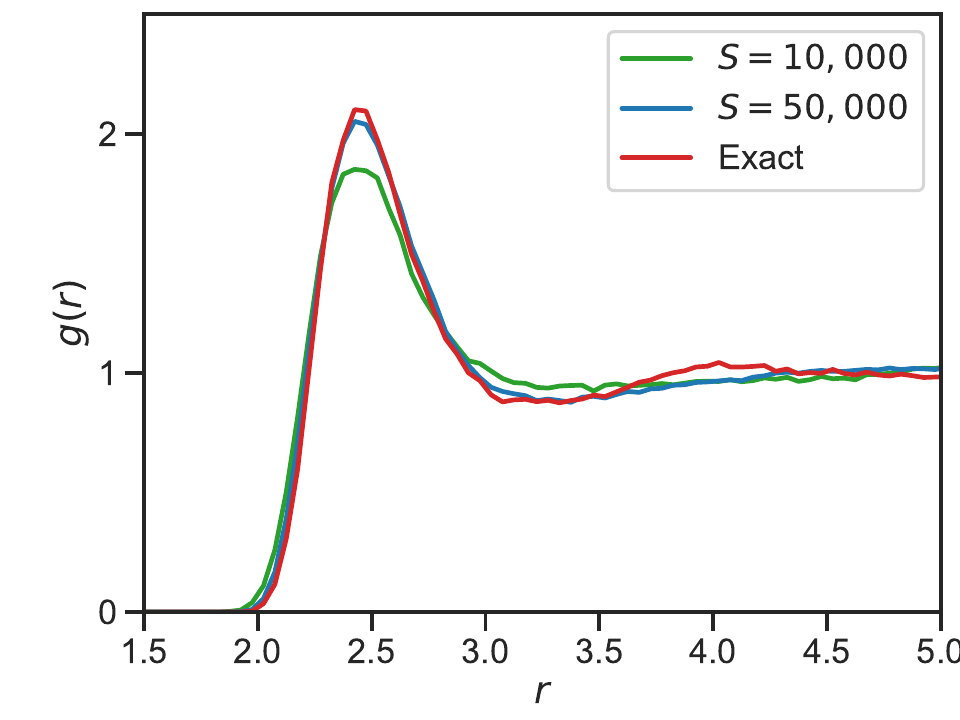}
\caption{Dependence of the radial distribution function of liquid silicon at $T=2000~\text{K}$ on the number of measurement shots $S$. To estimate the radial distribution function, we performed canonical MC simulations.}
\label{fig:MC_rdf}
\end{figure}

A critical consequence of using finite measurement shots is the difficulty in atomic force predictions. The stochastic nature of shot-based energy evaluations is incompatible with standard gradient methods. This renders the direct calculation of the analytic forces needed for MD simulations impractical in quantum hardware~\cite{obrien_efficient_2022}. To circumvent this limitation while still assessing the impact of shot noise on physical properties, we employed Monte Carlo (MC) simulations, which rely only on energy evaluations. We performed canonical MC simulations of liquid silicon at 2000~K to calculate the RDF (Figure~\ref{fig:MC_rdf}). The results clearly demonstrate that the calculated RDF systematically converges to the exact result (obtained without shot noise) as the number of measurement shots increases. This confirms that accurate structural properties can be recovered, but highlights the inherent trade-off between statistical accuracy and the computational cost of the simulation.

\subsection{Conclusion}

In this work, we developed and systematically evaluated a series of HQC-MLPs for the challenging task of simulating high-temperature (2000~K) liquid silicon. Our findings demonstrate that these models, which couple an $E(3)$-equivariant MPNN with a VQC, can accurately reproduce the underlying DFT potential energy surface. This was validated through stable canonical MD simulations, which successfully replicated key structural properties, such as the radial distribution function, in excellent agreement with our DFT reference.

Our investigation reveals a nuanced trade-off between training speed and generalization. A key advantage of the hybrid approach was a consistently accelerated convergence in training loss compared to the purely classical model. However, this was often accompanied by an earlier onset of overfitting. Crucially, we found that the model's ability to generalize is highly sensitive to the VQC architecture, with the choice of entangling gate being paramount. The models employing controlled-$Z$~(\texttt{CZ}) gates consistently outperformed all other variants, achieving the lowest validation loss and demonstrating a clear path to robust performance. This suggests that the specific nature of the entanglement operation is more critical than its mere presence.

Finally, we assessed the practical viability of these models by analyzing the impact of finite measurement shots. We confirmed that achieving chemical accuracy requires a substantial number of shots (on the order of $10^4-10^5$) and that the resulting statistical noise makes force calculation impractical, necessitating a shift to Monte Carlo methods for simulations.
Future work should therefore focus on two key areas: (1) the co-design of VQC architectures, leveraging the observed superiority of \texttt{CZ}-based entanglers to mitigate overfitting, and (2) the development of efficient algorithms to handle or bypass the challenges of shot noise in force estimation.
By providing a detailed map of both the advantages and current hurdles, this work helps pave the way for the practical application of hybrid quantum-classical machine learning models in atomistic simulations on quantum computers.

\section{Methods}

\subsection{Generation of training and validation datasets}

The dataset used to train the machine learning potential model was generated from BOMD simulations of liquid silicon at 3000~K.
This high-temperature regime was chosen to capture the complex dynamics and structural properties of liquid silicon~\cite{behlerGeneralizedNeuralNetworkRepresentation2007}, providing a challenging test case for our HQC-MLP models.
The system consisted of 64 silicon atoms in a periodic simulation cell with dimensions 10.862~\AA ~$\times$ 10.862~\AA ~$\times$ 10.862~\AA.
Silicon was modeled with the hybrid Gaussian and plane-wave (GPW) method~\cite{GPW1997}.
Core electrons were excluded by applying norm-conserving pseudopotentials developed by Goedecker, Teter, and Hutter~(GTH)~\cite{GTH1996,HGH1998}.
A charge density cutoff of 400~Ry was set for the auxiliary basis set.
The Perdew-Burke-Ernzerhof~(PBE) exchange-correlation functional~\cite{PBE1996} was used, and 
the Kohn-Sham orbitals were expanded into a double-$\zeta$ valence basis (denoted as DZVP) for the PBE functional.
To ensure strict convergence of the wavefunction, 
the electronic gradient convergence criterion was set to $\epsilon_\text{SCF} \le 5 \times 10^{-5}$, using the orbital transformation method to optimize single-determinant wave functions. 
BOMD simulations were performed using \textsc{Quickstep}~\cite{VANDEVONDELE2005103}, which is a part of the \textsc{CP2K} package~\cite{cp2k2020}.
The nuclear equation of motion was integrated using a velocity Verlet algorithm~\cite{velocityVerlet1982} with a time step of 0.5~fs, and
a Nos\'{e}-Hoover-chain thermostat~\cite{NHC1992} was employed to regulate the temperature.
These BOMD simulations generated a comprehensive set of atomic configurations, along with their corresponding energies and forces, which were crucial for training the model. 

\subsection{Generation of models: training and optimization processes}

The model architecture includes 5 interaction layers. 
The hidden feature dimension in each layer is represented in \texttt{e3nn} notation~\cite{e3nn_paper,e3nn} as ``64x0e + 64x0o + 64x1o + 64x1e + 8x2o + 8x2e''.
Radial basis vectors $\tilde{e}_{ij}^\text{RBF}$ in Eq.~\ref{eq:RBF} are generated using 8 radial Bessel basis functions and a polynomial envelope for the cutoff with $p=2$~\cite{gasteiger_directional_2021}.  
These radial features were then input into a multilayer perceptron $\mathcal{M}$ of size [64,~64], using \texttt{SiLU} nonlinearities on the outputs of the hidden layers as shown in Eq.~\ref{eq:conv}. 
We used a cutoff of 4.5~\AA, giving an average of 25 neighboring atoms per atom. 
The loss function used for training was
\begin{align}
    \mathcal{L} & = \frac{1}{B}\sum_{b}^B (\hat{E}_b - E_b)^2 + 
    \frac{\gamma_F}{3BN} \sum_{i = 1}^{B \cdot N} \sum_{\alpha=1}^3 \left(-\frac{\partial \hat{E}}{\partial r_{i,\alpha}} - F_{i,\alpha}\right)^2,
\end{align}
where $B$ denotes the number of batches and $N$ the number of atoms in the batch.
$E_b$ and $\hat{E}_b$ are the true and predicted energies for each batch, respectively,  
$F_{i,\alpha}$ the force on the atom $i$ in the direction $\alpha \in \{x, y, z\}$. 
The force loss weighting factor $\gamma_F$ was set to 100.

The models were optimized using the \textsc{AMSGrad}~\cite{reddi_convergence_2018} variant of the \textsc{Adam} optimizer~\cite{kingma_adam_2017} with default parameters $\beta_1=0.9$, $\beta_2=0.999$, and $\epsilon=10^{-8}$. 
We used a learning rate of 0.01 and a batch size of 20. 
The learning rate was reduced using an on-plateau scheduler based on the validation loss with a patience of 250 epochs and a decay factor of 0.9. 
We utilized an exponential moving average with weight 0.99 to smooth out the noisy parameter updates during training, leading to faster convergence.
The models were trained with single-precision floating point format~(\textbf{float32}) on an NVIDIA GeForce RTX 4090 GPU.
The typical training time was approximately two days.

\subsection{Validation of models} 

\textbf{Molecular dynamics simulations}: To validate the trained HQC-MLP parameters, we performed classical isothermal (\textit{NVT}) MD simulations of liquid silicon at $T = 2000~\text{K}$ using the ASE package~\cite{larsen_atomic_2017}. The integration of the equation of motion with a time step of 0.5 fs was propagated according to combined Nos{\'e}-Hoover and Parrinello-Rahman dynamics~\cite{melchionnaHooverNPTDynamics1993,melchionnaConstrainedSystemsStatistical2000,holianTimereversibleEquilibriumNonequilibrium1990}.

\textbf{Monte Carlo simulations}: 
To estimate stochastic potential energy under conditions relevant to real quantum computing, where atomic forces cannot be calculated, we performed Monte Carlo simulations of liquid silicon at $T = 2000~\text{K}$. 
New configurations were sampled using the the Metropolis-Hastings algorithm~\cite{MC}.

\section{Acknowledgments}
We are grateful to Luning Zhao, Joshua Goings, Torin Stetina, and Jungsang Kim for helpful discussions. CWM acknowledge the support provided by the National Research Foundation of Korea (NRF) grants funded by the Korean government (MSIT) (Grant No.~RS-2023-00283929, RS-2022-NR072058). SYW and CWM acknowledge the support provided by the National Research Foundation of Korea (NRF) grants funded by the Korean government (MSIT) (Grant No.~RS-2024-00407680). DCY acknowledge the support provided by the National Research Foundation of Korea (NRF) grants funded by the Korean government (MSIT) (Grant No.~RS-2023-00250313). This research was supported by `Quantum Information Science R\&D Ecosystem Creation' through the National Research Foundation of Korea(NRF) funded by the Korean government (Ministry of Science and ICT(MSIT))(No. 2020M3H3A1110365). The authors are grateful for the computational support from the Korea Institute of Science and Technology Information~(KISTI) with the \textsc{Nurion} cluster (KSC-2023-CRE-0059, KSC-2023-CRE-0311, KSC-2023-CRE-0355, KSC-2023-CRE-0454, KSC-2023-CRE-0472, KSC-2023-CRE-0502, KSC-2024-CRE-0358,KSC-2025-CRE-0122, KSC-2025-CRE-0125). Computational work for this research was also partially performed on the \textsc{Olaf} supercomputer supported by IBS Research Solution Center and on the GPU cluster supported by NIPA.

\section*{Supporting Information}
Supporting Information is available for this paper. 

\balance

\bibliographystyle{unsrtnat}
\bibliography{refs-qmlp}

\begin{thebibliography}{74}
\providecommand{\natexlab}[1]{#1}
\providecommand{\url}[1]{\texttt{#1}}
\expandafter\ifx\csname urlstyle\endcsname\relax
  \providecommand{\doi}[1]{doi: #1}\else
  \providecommand{\doi}{doi: \begingroup \urlstyle{rm}\Url}\fi

\bibitem[Biamonte et~al.(2017)Biamonte, Wittek, Pancotti, Rebentrost, Wiebe,
  and Lloyd]{biamonteQuantumMachineLearning2017}
Jacob Biamonte, Peter Wittek, Nicola Pancotti, Patrick Rebentrost, Nathan
  Wiebe, and Seth Lloyd.
\newblock Quantum machine learning.
\newblock \emph{Nature}, 549\penalty0 (7671):\penalty0 195--202, September
  2017.
\newblock ISSN 0028-0836, 1476-4687.
\newblock \doi{10.1038/nature23474}.

\bibitem[Sajjan et~al.(2022)Sajjan, Li, Selvarajan, Sureshbabu, Kale, Gupta,
  Singh, and Kais]{sajjanQuantumMachineLearning2022}
Manas Sajjan, Junxu Li, Raja Selvarajan, Shree~Hari Sureshbabu, Sumit~Suresh
  Kale, Rishabh Gupta, Vinit Singh, and Sabre Kais.
\newblock Quantum machine learning for chemistry and physics.
\newblock \emph{Chem. Soc. Rev.}, 51\penalty0 (15):\penalty0 6475--6573, 2022.
\newblock ISSN 0306-0012, 1460-4744.
\newblock \doi{10.1039/D2CS00203E}.

\bibitem[Zaman et~al.(2024)Zaman, Marchisio, Hanif, and
  Shafique]{zamanSurveyQuantumMachine2024}
Kamila Zaman, Alberto Marchisio, Muhammad~Abdullah Hanif, and Muhammad
  Shafique.
\newblock A {{Survey}} on {{Quantum Machine Learning}}: {{Current Trends}},
  {{Challenges}}, {{Opportunities}}, and the {{Road Ahead}}, July 2024.
\newblock URL \url{https://arxiv.org/abs/2310.10315}.

\bibitem[Zaman et~al.(2025)Zaman, Ahmed, Hanif, Marchisio, and
  Shafique]{zaman_comparative_2025}
Kamila Zaman, Tasnim Ahmed, Muhammad~Abdullah Hanif, Alberto Marchisio, and
  Muhammad Shafique.
\newblock A {Comparative} {Analysis} of {Hybrid}-{Quantum} {Classical} {Neural}
  {Networks}.
\newblock In Hamid~R. Arabnia, Masami Takata, Leonidas Deligiannidis, Pablo
  Rivas, Masahito Ohue, and Nobuaki Yasuo, editors, \emph{Grid, {Cloud}, and
  {Cluster} {Computing}; {Quantum} {Technologies}; and {Modeling}, {Simulation}
  and {Visualization} {Methods}}, pages 102--115, Cham, March 2025. Springer
  Nature Switzerland.
\newblock ISBN 978-3-031-85884-0.
\newblock \doi{10.1007/978-3-031-85884-0_9}.

\bibitem[Tilly et~al.(2022)Tilly, Chen, Cao, Picozzi, Setia, Li, Grant,
  Wossnig, Rungger, Booth, and
  Tennyson]{tillyVariationalQuantumEigensolver2022}
Jules Tilly, Hongxiang Chen, Shuxiang Cao, Dario Picozzi, Kanav Setia, Ying Li,
  Edward Grant, Leonard Wossnig, Ivan Rungger, George~H. Booth, and Jonathan
  Tennyson.
\newblock The {{Variational Quantum Eigensolver}}: A review of methods and best
  practices.
\newblock \emph{Phys. Rep.}, 986:\penalty0 1--128, November 2022.
\newblock ISSN 03701573.
\newblock \doi{10.1016/j.physrep.2022.08.003}.

\bibitem[Huang et~al.(2023)Huang, Xu, Guo, Tian, Wei, Sun, Bao, and
  Long]{huangTermQuantumComputing2023}
He-Liang Huang, Xiao-Yue Xu, Chu Guo, Guojing Tian, Shi-Jie Wei, Xiaoming Sun,
  Wan-Su Bao, and Gui-Lu Long.
\newblock Near-term quantum computing techniques: {{Variational}} quantum
  algorithms, error mitigation, circuit compilation, benchmarking and classical
  simulation.
\newblock \emph{Sci. China Phys. Mech. Astron.}, 66\penalty0 (5):\penalty0
  250302, May 2023.
\newblock ISSN 1674-7348, 1869-1927.
\newblock \doi{10.1007/s11433-022-2057-y}.

\bibitem[Mari et~al.(2020)Mari, Bromley, Izaac, Schuld, and
  Killoran]{mariTransferLearningHybrid2020}
Andrea Mari, Thomas~R. Bromley, Josh Izaac, Maria Schuld, and Nathan Killoran.
\newblock Transfer learning in hybrid classical-quantum neural networks.
\newblock \emph{Quantum}, 4:\penalty0 340, October 2020.
\newblock ISSN 2521-327X.
\newblock \doi{10.22331/q-2020-10-09-340}.

\bibitem[Bowles et~al.(2024)Bowles, Ahmed, and
  Schuld]{bowlesBetterClassicalSubtle2024}
Joseph Bowles, Shahnawaz Ahmed, and Maria Schuld.
\newblock Better than classical? {{The}} subtle art of benchmarking quantum
  machine learning models, March 2024.
\newblock URL \url{https://arxiv.org/abs/2403.07059}.

\bibitem[West et~al.(2024)West, Heredge, Sevior, and Usman]{West2024}
Maxwell~T. West, Jamie Heredge, Martin Sevior, and Muhammad Usman.
\newblock Provably trainable rotationally equivariant quantum machine learning.
\newblock \emph{PRX Quantum}, 5:\penalty0 030320, Jul 2024.
\newblock \doi{10.1103/PRXQuantum.5.030320}.
\newblock URL \url{https://link.aps.org/doi/10.1103/PRXQuantum.5.030320}.

\bibitem[Thiemann et~al.(2025)Thiemann, O'Neill, Kapil, Michaelides, and
  Schran]{thiemann_introduction_2025_review}
Fabian~L Thiemann, Niamh O'Neill, Venkat Kapil, Angelos Michaelides, and
  Christoph Schran.
\newblock Introduction to machine learning potentials for atomistic
  simulations.
\newblock \emph{J. Phys.: Condens. Matter}, 37\penalty0 (7):\penalty0 073002,
  February 2025.
\newblock ISSN 0953-8984.
\newblock \doi{10.1088/1361-648X/ad9657}.
\newblock URL
  \url{https://iopscience.iop.org/article/10.1088/1361-648X/ad9657}.

\bibitem[Aldossary et~al.(2024)Aldossary, Campos-Gonzalez-Angulo,
  Pablo-García, Leong, Rajaonson, Thiede, Tom, Wang, Avagliano, and
  Aspuru-Guzik]{aldossary_silico_2024_review}
Abdulrahman Aldossary, Jorge~Arturo Campos-Gonzalez-Angulo, Sergio
  Pablo-García, Shi~Xuan Leong, Ella~Miray Rajaonson, Luca Thiede, Gary Tom,
  Andrew Wang, Davide Avagliano, and Alán Aspuru-Guzik.
\newblock In {Silico} {Chemical} {Experiments} in the {Age} of {AI}: {From}
  {Quantum} {Chemistry} to {Machine} {Learning} and {Back}.
\newblock \emph{Adv. Mater.}, 36\penalty0 (30):\penalty0 2402369, July 2024.
\newblock ISSN 1521-4095.
\newblock \doi{10.1002/adma.202402369}.
\newblock URL
  \url{https://onlinelibrary.wiley.com/doi/abs/10.1002/adma.202402369}.

\bibitem[Wan et~al.(2024)Wan, He, and Shi]{wan_construction_2024_review}
Kaiwei Wan, Jianxin He, and Xinghua Shi.
\newblock Construction of {High} {Accuracy} {Machine} {Learning} {Interatomic}
  {Potential} for {Surface}/{Interface} of {Nanomaterials}---{A} {Review}.
\newblock \emph{Adv. Mater.}, 36\penalty0 (22):\penalty0 2305758, May 2024.
\newblock ISSN 1521-4095.
\newblock \doi{10.1002/adma.202305758}.
\newblock URL
  \url{https://onlinelibrary.wiley.com/doi/abs/10.1002/adma.202305758}.

\bibitem[Kocer et~al.(2022)Kocer, Ko, and Behler]{kocer_neural_2022_review}
Emir Kocer, Tsz~Wai Ko, and Jörg Behler.
\newblock Neural {Network} {Potentials}: {A} {Concise} {Overview} of {Methods}.
\newblock \emph{Annu. Rev. Phys. Chem.}, 73\penalty0 (1):\penalty0 163--186,
  April 2022.
\newblock \doi{10.1146/annurev-physchem-082720-034254}.
\newblock URL
  \url{https://www.annualreviews.org/doi/abs/10.1146/annurev-physchem-082720-034254}.

\bibitem[Unke et~al.(2021)Unke, Chmiela, Sauceda, Gastegger, Poltavsky,
  Sch{\"u}tt, Tkatchenko, and M{\"u}ller]{unke_mlff_2021_review}
Oliver~T. Unke, Stefan Chmiela, Huziel~E. Sauceda, Michael Gastegger, Igor
  Poltavsky, Kristof~T. Sch{\"u}tt, Alexandre Tkatchenko, and Klaus-Robert
  M{\"u}ller.
\newblock Machine {{Learning Force Fields}}.
\newblock \emph{Chem. Rev.}, 121\penalty0 (16):\penalty0 10142--10186, August
  2021.
\newblock ISSN 0009-2665, 1520-6890.
\newblock \doi{10.1021/acs.chemrev.0c01111}.

\bibitem[Behler(2021)]{behler_four_2021_review}
Jörg Behler.
\newblock Four {Generations} of {High}-{Dimensional} {Neural} {Network}
  {Potentials}.
\newblock \emph{Chem. Rev.}, 121\penalty0 (16):\penalty0 10037--10072, August
  2021.
\newblock ISSN 0009-2665.
\newblock \doi{10.1021/acs.chemrev.0c00868}.
\newblock URL \url{https://pubs.acs.org/doi/10.1021/acs.chemrev.0c00868}.

\bibitem[Dral(2020)]{dral_quantum_2020_review}
Pavlo~O. Dral.
\newblock Quantum {Chemistry} in the {Age} of {Machine} {Learning}.
\newblock \emph{J. Phys. Chem. Lett.}, 11\penalty0 (6):\penalty0 2336--2347,
  March 2020.
\newblock \doi{10.1021/acs.jpclett.9b03664}.
\newblock URL \url{https://doi.org/10.1021/acs.jpclett.9b03664}.

\bibitem[Merchant et~al.(2023)Merchant, Batzner, Schoenholz, Aykol, Cheon, and
  Cubuk]{merchant_scaling_2023}
Amil Merchant, Simon Batzner, Samuel~S. Schoenholz, Muratahan Aykol, Gowoon
  Cheon, and Ekin~Dogus Cubuk.
\newblock Scaling deep learning for materials discovery.
\newblock \emph{Nature}, 624\penalty0 (7990):\penalty0 80--85, December 2023.
\newblock ISSN 0028-0836, 1476-4687.
\newblock \doi{10.1038/s41586-023-06735-9}.

\bibitem[Batatia et~al.(2024)Batatia, Benner, Chiang, Elena, Kov{\'a}cs,
  Riebesell, Advincula, Asta, Avaylon, Baldwin, Berger, Bernstein, Bhowmik,
  Blau, C{\u a}rare, Darby, De, Della~Pia, Deringer, Elijo{\v s}ius,
  {El-Machachi}, Falcioni, Fako, Ferrari, {Genreith-Schriever}, George,
  Goodall, Grey, Grigorev, Han, Handley, Heenen, Hermansson, Holm, Jaafar,
  Hofmann, Jakob, Jung, Kapil, Kaplan, Karimitari, Kermode, Kroupa, Kullgren,
  Kuner, Kuryla, Liepuoniute, Margraf, Magd{\u a}u, Michaelides, Moore, Naik,
  Niblett, Norwood, O'Neill, Ortner, Persson, Reuter, Rosen, Schaaf, Schran,
  Shi, Sivonxay, Stenczel, Svahn, Sutton, Swinburne, Tilly, {van der Oord},
  {Varga-Umbrich}, Vegge, Vondr{\'a}k, Wang, Witt, Zills, and
  Cs{\'a}nyi]{batatia_foundation_2024}
Ilyes Batatia, Philipp Benner, Yuan Chiang, Alin~M. Elena, D{\'a}vid~P.
  Kov{\'a}cs, Janosh Riebesell, Xavier~R. Advincula, Mark Asta, Matthew
  Avaylon, William~J. Baldwin, Fabian Berger, Noam Bernstein, Arghya Bhowmik,
  Samuel~M. Blau, Vlad C{\u a}rare, James~P. Darby, Sandip De, Flaviano
  Della~Pia, Volker~L. Deringer, Rokas Elijo{\v s}ius, Zakariya {El-Machachi},
  Fabio Falcioni, Edvin Fako, Andrea~C. Ferrari, Annalena {Genreith-Schriever},
  Janine George, Rhys E.~A. Goodall, Clare~P. Grey, Petr Grigorev, Shuang Han,
  Will Handley, Hendrik~H. Heenen, Kersti Hermansson, Christian Holm, Jad
  Jaafar, Stephan Hofmann, Konstantin~S. Jakob, Hyunwook Jung, Venkat Kapil,
  Aaron~D. Kaplan, Nima Karimitari, James~R. Kermode, Namu Kroupa, Jolla
  Kullgren, Matthew~C. Kuner, Domantas Kuryla, Guoda Liepuoniute, Johannes~T.
  Margraf, Ioan-Bogdan Magd{\u a}u, Angelos Michaelides, J.~Harry Moore,
  Aakash~A. Naik, Samuel~P. Niblett, Sam~Walton Norwood, Niamh O'Neill,
  Christoph Ortner, Kristin~A. Persson, Karsten Reuter, Andrew~S. Rosen,
  Lars~L. Schaaf, Christoph Schran, Benjamin~X. Shi, Eric Sivonxay,
  Tam{\'a}s~K. Stenczel, Viktor Svahn, Christopher Sutton, Thomas~D. Swinburne,
  Jules Tilly, Cas {van der Oord}, Eszter {Varga-Umbrich}, Tejs Vegge, Martin
  Vondr{\'a}k, Yangshuai Wang, William~C. Witt, Fabian Zills, and G{\'a}bor
  Cs{\'a}nyi.
\newblock A foundation model for atomistic materials chemistry, March 2024.
\newblock URL \url{https://arxiv.org/abs/2401.00096}.

\bibitem[Myung et~al.(2022)Myung, Hajibabaei, Cha, Ha, Kim, and
  Kim]{myung_challenges_2022}
Chang~Woo Myung, Amir Hajibabaei, Ji-Hyun Cha, Miran Ha, Junu Kim, and Kwang~S.
  Kim.
\newblock Challenges, {{Opportunities}}, and {{Prospects}} in {{Metal Halide
  Perovskites}} from {{Theoretical}} and {{Machine Learning Perspectives}}.
\newblock \emph{Adv. Energy Mater.}, 12\penalty0 (45):\penalty0 2202279,
  December 2022.
\newblock ISSN 1614-6832, 1614-6840.
\newblock \doi{10.1002/aenm.202202279}.

\bibitem[Behler and
  Parrinello(2007)]{behlerGeneralizedNeuralNetworkRepresentation2007}
J{\"o}rg Behler and Michele Parrinello.
\newblock Generalized {{Neural-Network Representation}} of {{High-Dimensional
  Potential-Energy Surfaces}}.
\newblock \emph{Phys. Rev. Lett.}, 98\penalty0 (14):\penalty0 146401, April
  2007.
\newblock ISSN 0031-9007, 1079-7114.
\newblock \doi{10.1103/PhysRevLett.98.146401}.

\bibitem[Eckhoff and Behler(2021)]{eckhoffHighdimensionalNeuralNetwork2021}
Marco Eckhoff and J{\"o}rg Behler.
\newblock High-dimensional neural network potentials for magnetic systems using
  spin-dependent atom-centered symmetry functions.
\newblock \emph{npj Comput. Mater.}, 7\penalty0 (1):\penalty0 170, December
  2021.
\newblock ISSN 2057-3960.
\newblock \doi{10.1038/s41524-021-00636-z}.

\bibitem[Bart{\'o}k et~al.(2010)Bart{\'o}k, Payne, Kondor, and
  Cs{\'a}nyi]{bartok_gap_2010}
Albert~P. Bart{\'o}k, Mike~C. Payne, Risi Kondor, and G{\'a}bor Cs{\'a}nyi.
\newblock Gaussian {{Approximation Potentials}}: {{The Accuracy}} of {{Quantum
  Mechanics}}, without the {{Electrons}}.
\newblock \emph{Phys. Rev. Lett.}, 104\penalty0 (13):\penalty0 136403, April
  2010.
\newblock ISSN 0031-9007, 1079-7114.
\newblock \doi{10.1103/PhysRevLett.104.136403}.

\bibitem[Bart{\'o}k et~al.(2013)Bart{\'o}k, Kondor, and
  Cs{\'a}nyi]{bartok_representing_2013}
Albert~P. Bart{\'o}k, Risi Kondor, and G{\'a}bor Cs{\'a}nyi.
\newblock On representing chemical environments.
\newblock \emph{Phys. Rev. B}, 87\penalty0 (18):\penalty0 184115, May 2013.
\newblock ISSN 1098-0121, 1550-235X.
\newblock \doi{10.1103/PhysRevB.87.184115}.

\bibitem[Bartók and Csányi(2015)]{bartok_gaussian_2015}
Albert~P. Bartók and Gábor Csányi.
\newblock Gaussian approximation potentials: {A} brief tutorial introduction.
\newblock \emph{Int. J. Quantum Chem.}, 115\penalty0 (16):\penalty0 1051--1057,
  August 2015.
\newblock ISSN 1097-461X.
\newblock \doi{10.1002/qua.24927}.
\newblock URL \url{https://onlinelibrary.wiley.com/doi/10.1002/qua.24927}.

\bibitem[Klawohn et~al.(2023)Klawohn, Darby, Kermode, Cs{\'a}nyi, Caro, and
  Bart{\'o}k]{klawohn_gap_2023}
Sascha Klawohn, James~P. Darby, James~R. Kermode, G{\'a}bor Cs{\'a}nyi,
  Miguel~A. Caro, and Albert~P. Bart{\'o}k.
\newblock Gaussian approximation potentials: {{Theory}}, software
  implementation and application examples.
\newblock \emph{J. Chem. Phys.}, 159\penalty0 (17):\penalty0 174108, November
  2023.
\newblock ISSN 0021-9606, 1089-7690.
\newblock \doi{10.1063/5.0160898}.

\bibitem[Hajibabaei et~al.(2021{\natexlab{a}})Hajibabaei, Ha, Pourasad, Kim,
  and Kim]{hajibabaei_machine_2021}
Amir Hajibabaei, Miran Ha, Saeed Pourasad, Junu Kim, and Kwang~S. Kim.
\newblock Machine {{Learning}} of {{First-Principles Force-Fields}} for
  {{Alkane}} and {{Polyene Hydrocarbons}}.
\newblock \emph{J. Phys. Chem. A}, 125\penalty0 (42):\penalty0 9414--9420,
  October 2021{\natexlab{a}}.
\newblock ISSN 1089-5639, 1520-5215.
\newblock \doi{10.1021/acs.jpca.1c05819}.

\bibitem[Hajibabaei et~al.(2021{\natexlab{b}})Hajibabaei, Myung, and
  Kim]{hajibabaei_sparse_2021}
Amir Hajibabaei, Chang~Woo Myung, and Kwang~S. Kim.
\newblock Sparse {{Gaussian}} process potentials: {{Application}} to lithium
  diffusivity in superionic conducting solid electrolytes.
\newblock \emph{Phys. Rev. B}, 103\penalty0 (21):\penalty0 214102, June
  2021{\natexlab{b}}.
\newblock ISSN 2469-9950, 2469-9969.
\newblock \doi{10.1103/PhysRevB.103.214102}.

\bibitem[Hajibabaei and Kim(2021)]{hajibabaei_universal_2021}
Amir Hajibabaei and Kwang~S. Kim.
\newblock Universal {{Machine Learning Interatomic Potentials}}: {{Surveying
  Solid Electrolytes}}.
\newblock \emph{J. Phys. Chem. Lett.}, 12\penalty0 (33):\penalty0 8115--8120,
  August 2021.
\newblock ISSN 1948-7185, 1948-7185.
\newblock \doi{10.1021/acs.jpclett.1c01605}.

\bibitem[Ha et~al.(2022)Ha, Hajibabaei, Pourasad, and Kim]{ha_sparse_2022}
Miran Ha, Amir Hajibabaei, Saeed Pourasad, and Kwang~S. Kim.
\newblock Sparse {{Gaussian Process Regression-Based Machine Learned
  First-Principles Force-Fields}} for {{Saturated}}, {{Olefinic}}, and
  {{Aromatic Hydrocarbons}}.
\newblock \emph{ACS Phys. Chem Au}, 2\penalty0 (3):\penalty0 260--264, May
  2022.
\newblock ISSN 2694-2445, 2694-2445.
\newblock \doi{10.1021/acsphyschemau.1c00058}.

\bibitem[Willow et~al.(2024)Willow, Kim, Sundheep, Hajibabaei, Kim, and
  Myung]{willow_active_2024}
Soohaeng~Yoo Willow, Dong~Geon Kim, R.~Sundheep, Amir Hajibabaei, Kwang~S. Kim,
  and Chang~Woo Myung.
\newblock Active sparse {{Bayesian}} committee machine potential for
  isothermal--isobaric molecular dynamics simulations.
\newblock \emph{Phys. Chem. Chem. Phys.}, 26\penalty0 (33):\penalty0
  22073--22082, 2024.
\newblock ISSN 1463-9076, 1463-9084.
\newblock \doi{10.1039/D4CP01801J}.

\bibitem[{Qui{\~n}onero-Candela} and
  Rasmussen(2005)]{quinonero-candela_unifying_2005}
Joaquin {Qui{\~n}onero-Candela} and Carl~Edward Rasmussen.
\newblock A {{Unifying View}} of {{Sparse Approximate Gaussian Process
  Regression}}.
\newblock \emph{J. Mach. Learn. Res.}, 6\penalty0 (65):\penalty0 1939--1959,
  2005.

\bibitem[Geiger and Smidt(2022)]{e3nn_paper}
Mario Geiger and Tess Smidt.
\newblock e3nn: Euclidean neural networks.
\newblock \emph{arXiv e-prints}, page arXiv:2207.09453, 2022.
\newblock URL \url{https://arxiv.org/abs/2207.09453}.

\bibitem[Batzner et~al.(2022)Batzner, Musaelian, Sun, Geiger, Mailoa,
  Kornbluth, Molinari, Smidt, and Kozinsky]{batzner_e3-equivariant_2022}
Simon Batzner, Albert Musaelian, Lixin Sun, Mario Geiger, Jonathan~P. Mailoa,
  Mordechai Kornbluth, Nicola Molinari, Tess~E. Smidt, and Boris Kozinsky.
\newblock E(3)-equivariant graph neural networks for data-efficient and
  accurate interatomic potentials.
\newblock \emph{Nat. Commun.}, 13\penalty0 (1):\penalty0 2453, May 2022.
\newblock ISSN 2041-1723.
\newblock \doi{10.1038/s41467-022-29939-5}.

\bibitem[Cohen et~al.(2019{\natexlab{a}})Cohen, Geiger, and
  Weiler]{cohen_general_2019}
Taco~S. Cohen, Mario Geiger, and Maurice Weiler.
\newblock A general theory of equivariant {CNNs} on homogeneous spaces.
\newblock In \emph{Proceedings of the 33rd {International} {Conference} on
  {Neural} {Information} {Processing} {Systems}}, {NIPS} '19, pages 9145--9156,
  Red Hook, NY, USA, December 2019{\natexlab{a}}. Curran Associates Inc.
\newblock \doi{10.5555/3454287.3455107}.
\newblock URL \url{https://dl.acm.org/doi/10.5555/3454287.3455107}.
\newblock
  https://papers.nips.cc/paper\_files/paper/2019/hash/b9cfe8b6042cf759dc4c0cccb27a6737-Abstract.html.

\bibitem[Cohen et~al.(2019{\natexlab{b}})Cohen, Weiler, Kicanaoglu, and
  Welling]{cohen_gauge_2019}
Taco Cohen, Maurice Weiler, Berkay Kicanaoglu, and Max Welling.
\newblock Gauge {Equivariant} {Convolutional} {Networks} and the {Icosahedral}
  {CNN}.
\newblock In \emph{Proceedings of the 36th {International} {Conference} on
  {Machine} {Learning}}, volume~97, pages 1321--1330, Long Beach, CA, USA, May
  2019{\natexlab{b}}. ML Research Press.
\newblock URL \url{https://proceedings.mlr.press/v97/cohen19d.html}.
\newblock ISSN: 2640-3498.

\bibitem[Cerezo et~al.(2022)Cerezo, Verdon, Huang, Cincio, and
  Coles]{cerezo_challenges_2022}
M.~Cerezo, Guillaume Verdon, Hsin-Yuan Huang, Lukasz Cincio, and Patrick~J.
  Coles.
\newblock Challenges and opportunities in quantum machine learning.
\newblock \emph{Nat. Comput. Sci.}, 2\penalty0 (9):\penalty0 567--576,
  September 2022.
\newblock ISSN 2662-8457.
\newblock \doi{10.1038/s43588-022-00311-3}.
\newblock URL \url{https://doi.org/10.1038/s43588-022-00311-3}.

\bibitem[Shiota et~al.(2024)Shiota, Ishihara, and
  Mizukami]{shiota_universal_2024}
Tomoya Shiota, Kenji Ishihara, and Wataru Mizukami.
\newblock Universal neural network potentials as descriptors: towards scalable
  chemical property prediction using quantum and classical computers.
\newblock \emph{Digit. Discov.}, 3\penalty0 (9):\penalty0 1714--1728, September
  2024.
\newblock ISSN 2635-098X.
\newblock \doi{10.1039/D4DD00098F}.
\newblock URL
  \url{https://pubs.rsc.org/en/content/articlelanding/2024/dd/d4dd00098f}.

\bibitem[Couzinié et~al.(2025)Couzinié, Daimon, Nishi, Ito, Harazono, and
  Matsushita]{couzinie_towards_2025}
Yannick Couzinié, Shunsuke Daimon, Hirofumi Nishi, Natsuki Ito, Yusuke
  Harazono, and Yu-ichiro Matsushita.
\newblock Towards {Improved} {Quantum} {Machine} {Learning} for {Molecular}
  {Force} {Fields}, May 2025.
\newblock URL \url{https://arxiv.org/abs/2505.03213}.

\bibitem[Le~Gall(2025)]{le_gall_robust_2025}
François Le~Gall.
\newblock Robust {Dequantization} of the {Quantum} {Singular} {Value}
  {Transformation} and {Quantum} {Machine} {Learning} {Algorithms}.
\newblock \emph{comput. complex.}, 34\penalty0 (1):\penalty0 2, January 2025.
\newblock ISSN 1420-8954.
\newblock \doi{10.1007/s00037-024-00262-3}.
\newblock URL
  \url{https://link.springer.com/article/10.1007/s00037-024-00262-3}.

\bibitem[Chia et~al.(2020)Chia, Gilyén, Li, Lin, Tang, and
  Wang]{chia_sampling-based_2020}
Nai-Hui Chia, András Gilyén, Tongyang Li, Han-Hsuan Lin, Ewin Tang, and
  Chunhao Wang.
\newblock Sampling-based sublinear low-rank matrix arithmetic framework for
  dequantizing {Quantum} machine learning.
\newblock In \emph{Proceedings of the 52nd {Annual} {ACM} {SIGACT} {Symposium}
  on {Theory} of {Computing}}, pages 387--400, Chicago IL USA, June 2020. ACM.
\newblock ISBN 978-1-4503-6979-4.
\newblock \doi{10.1145/3357713.3384314}.
\newblock URL \url{https://dl.acm.org/doi/10.1145/3357713.3384314}.

\bibitem[Babbush et~al.(2021)Babbush, McClean, Newman, Gidney, Boixo, and
  Neven]{babbush_focus_2021}
Ryan Babbush, Jarrod~R. McClean, Michael Newman, Craig Gidney, Sergio Boixo,
  and Hartmut Neven.
\newblock Focus beyond {Quadratic} {Speedups} for {Error}-{Corrected} {Quantum}
  {Advantage}.
\newblock \emph{PRX Quantum}, 2\penalty0 (1):\penalty0 010103, March 2021.
\newblock ISSN 2691-3399.
\newblock \doi{10.1103/PRXQuantum.2.010103}.
\newblock URL \url{https://link.aps.org/doi/10.1103/PRXQuantum.2.010103}.

\bibitem[Grover(1996)]{grover_fast_1996}
Lov~K. Grover.
\newblock A fast quantum mechanical algorithm for database search.
\newblock In \emph{Proceedings of the twenty-eighth annual {ACM} symposium on
  {Theory} of computing - {STOC} '96}, pages 212--219, Philadelphia,
  Pennsylvania, United States, 1996. ACM Press.
\newblock ISBN 978-0-89791-785-8.
\newblock \doi{10.1145/237814.237866}.
\newblock URL \url{http://portal.acm.org/citation.cfm?doid=237814.237866}.

\bibitem[Bernstein and Vazirani(1997)]{Bernstein_1997}
Ethan Bernstein and Umesh Vazirani.
\newblock Quantum complexity theory.
\newblock \emph{SIAM Journal on Computing}, 26\penalty0 (5):\penalty0
  1411--1473, 1997.
\newblock \doi{10.1137/S0097539796300921}.
\newblock URL \url{https://doi.org/10.1137/S0097539796300921}.

\bibitem[Zi et~al.(2024)Zi, Wang, Kim, Sun, Chattopadhyay, and
  Rebentrost]{zi_efficient_2024}
Wei Zi, Siyi Wang, Hyunji Kim, Xiaoming Sun, Anupam Chattopadhyay, and Patrick
  Rebentrost.
\newblock Efficient quantum circuits for machine learning activation functions
  including constant {T}-depth {ReLU}.
\newblock 6\penalty0 (4):\penalty0 043048, October 2024.
\newblock \doi{10.1103/PhysRevResearch.6.043048}.
\newblock URL
  \url{https://journals.aps.org/prresearch/abstract/10.1103/PhysRevResearch.6.043048}.

\bibitem[Stillinger and Weber(1985)]{stillinger_computer_1985}
Frank~H. Stillinger and Thomas~A. Weber.
\newblock Computer simulation of local order in condensed phases of silicon.
\newblock \emph{Phys. Rev. B}, 31\penalty0 (8):\penalty0 5262--5271, April
  1985.
\newblock \doi{10.1103/PhysRevB.31.5262}.
\newblock URL \url{https://link.aps.org/doi/10.1103/PhysRevB.31.5262}.

\bibitem[Lenosky et~al.(2000)Lenosky, Sadigh, Alonso, Bulatov, de~la Rubia,
  Kim, Voter, and Kress]{Lenosky_2000}
Thomas~J Lenosky, Babak Sadigh, Eduardo Alonso, Vasily~V Bulatov, Tomas~Diaz
  de~la Rubia, Jeongnim Kim, Arthur~F Voter, and Joel~D Kress.
\newblock Highly optimized empirical potential model of silicon.
\newblock \emph{Model. Simul. Mater. Sci. Eng.}, 8\penalty0 (6):\penalty0 825,
  nov 2000.
\newblock \doi{10.1088/0965-0393/8/6/305}.
\newblock URL \url{https://dx.doi.org/10.1088/0965-0393/8/6/305}.

\bibitem[Zhang et~al.(2015)Zhang, Wang, Zhang, Qi, Zhang, Ma, and
  Liu]{zhang_polymorphism_2015}
Shiliang Zhang, Li-Min Wang, Xinyu Zhang, Li~Qi, Suhong Zhang, Mingzhen Ma, and
  Riping Liu.
\newblock Polymorphism in glassy silicon: {Inherited} from liquid-liquid phase
  transition in supercooled liquid.
\newblock \emph{Sci. Rep.}, 5\penalty0 (1):\penalty0 8590, February 2015.
\newblock ISSN 2045-2322.
\newblock \doi{10.1038/srep08590}.
\newblock URL \url{https://www.nature.com/articles/srep08590}.

\bibitem[Gasteiger et~al.(2021)Gasteiger, Yeshwanth, and
  Günnemann]{gasteiger_directional_2021}
Johannes Gasteiger, Chandan Yeshwanth, and Stephan Günnemann.
\newblock Directional {Message} {Passing} on {Molecular} {Graphs} via
  {Synthetic} {Coordinates}.
\newblock In \emph{Advances in {Neural} {Information} {Processing} {Systems}},
  volume~34, pages 15421--15433. Curran Associates, Inc., December 2021.
\newblock URL
  \url{https://proceedings.neurips.cc/paper/2021/hash/82489c9737cc245530c7a6ebef3753ec-Abstract.html}.

\bibitem[He et~al.(2016)He, Zhang, Ren, and Sun]{he_deep_2016}
Kaiming He, Xiangyu Zhang, Shaoqing Ren, and Jian Sun.
\newblock Deep {Residual} {Learning} for {Image} {Recognition}.
\newblock In \emph{Proceedings of the {IEEE} {Conference} on {Computer}
  {Vision} and {Pattern} {Recognition}}, volume 2016, pages 770--778, June
  2016.
\newblock URL
  \url{https://openaccess.thecvf.com/content_cvpr_2016/html/He_Deep_Residual_Learning_CVPR_2016_paper.html}.

\bibitem[Bergholm et~al.(2022)Bergholm, Izaac, Schuld, Gogolin, Ahmed, Ajith,
  Alam, {Alonso-Linaje}, AkashNarayanan, Asadi, Arrazola, Azad, Banning, Blank,
  Bromley, Cordier, Ceroni, Delgado, Di~Matteo, Dusko, Garg, Guala, Hayes,
  Hill, Ijaz, Isacsson, Ittah, Jahangiri, Jain, Jiang, Khandelwal, Kottmann,
  Lang, Lee, Loke, Lowe, McKiernan, Meyer, {Monta{\~n}ez-Barrera}, Moyard, Niu,
  O'Riordan, Oud, Panigrahi, Park, Polatajko, Quesada, Roberts, S{\'a}, Schoch,
  Shi, Shu, Sim, Singh, Strandberg, Soni, Sz{\'a}va, Thabet,
  {Vargas-Hern{\'a}ndez}, Vincent, Vitucci, Weber, Wierichs, Wiersema,
  Willmann, Wong, Zhang, and
  Killoran]{bergholmPennyLaneAutomaticDifferentiation2022}
Ville Bergholm, Josh Izaac, Maria Schuld, Christian Gogolin, Shahnawaz Ahmed,
  Vishnu Ajith, M.~Sohaib Alam, Guillermo {Alonso-Linaje}, B.~AkashNarayanan,
  Ali Asadi, Juan~Miguel Arrazola, Utkarsh Azad, Sam Banning, Carsten Blank,
  Thomas~R. Bromley, Benjamin~A. Cordier, Jack Ceroni, Alain Delgado, Olivia
  Di~Matteo, Amintor Dusko, Tanya Garg, Diego Guala, Anthony Hayes, Ryan Hill,
  Aroosa Ijaz, Theodor Isacsson, David Ittah, Soran Jahangiri, Prateek Jain,
  Edward Jiang, Ankit Khandelwal, Korbinian Kottmann, Robert~A. Lang, Christina
  Lee, Thomas Loke, Angus Lowe, Keri McKiernan, Johannes~Jakob Meyer, J.~A.
  {Monta{\~n}ez-Barrera}, Romain Moyard, Zeyue Niu, Lee~James O'Riordan, Steven
  Oud, Ashish Panigrahi, Chae-Yeun Park, Daniel Polatajko, Nicol{\'a}s Quesada,
  Chase Roberts, Nahum S{\'a}, Isidor Schoch, Borun Shi, Shuli Shu, Sukin Sim,
  Arshpreet Singh, Ingrid Strandberg, Jay Soni, Antal Sz{\'a}va, Slimane
  Thabet, Rodrigo~A. {Vargas-Hern{\'a}ndez}, Trevor Vincent, Nicola Vitucci,
  Maurice Weber, David Wierichs, Roeland Wiersema, Moritz Willmann, Vincent
  Wong, Shaoming Zhang, and Nathan Killoran.
\newblock {{PennyLane}}: {{Automatic}} differentiation of hybrid
  quantum-classical computations, July 2022.

\bibitem[McClean et~al.(2018)McClean, Boixo, Smelyanskiy, Babbush, and
  Neven]{mcclean_barren_2018}
Jarrod~R. McClean, Sergio Boixo, Vadim~N. Smelyanskiy, Ryan Babbush, and
  Hartmut Neven.
\newblock Barren plateaus in quantum neural network training landscapes.
\newblock \emph{Nat. Commun.}, 9\penalty0 (1):\penalty0 4812, November 2018.
\newblock ISSN 2041-1723.
\newblock \doi{10.1038/s41467-018-07090-4}.
\newblock URL \url{https://www.nature.com/articles/s41467-018-07090-4}.

\bibitem[Holmes et~al.(2022)Holmes, Sharma, Cerezo, and
  Coles]{Holmes2022Connecting}
Zo\"e Holmes, Kunal Sharma, M.~Cerezo, and Patrick~J. Coles.
\newblock Connecting ansatz expressibility to gradient magnitudes and barren
  plateaus.
\newblock \emph{PRX Quantum}, 3:\penalty0 010313, Jan 2022.
\newblock \doi{10.1103/PRXQuantum.3.010313}.
\newblock URL \url{https://link.aps.org/doi/10.1103/PRXQuantum.3.010313}.

\bibitem[Cerezo et~al.(2021)Cerezo, Sone, Volkoff, Cincio, and
  Coles]{cerezo_cost_2021}
M.~Cerezo, Akira Sone, Tyler Volkoff, Lukasz Cincio, and Patrick~J. Coles.
\newblock Cost function dependent barren plateaus in shallow parametrized
  quantum circuits.
\newblock \emph{Nat. Commun.}, 12\penalty0 (1):\penalty0 1791, March 2021.
\newblock ISSN 2041-1723.
\newblock \doi{10.1038/s41467-021-21728-w}.
\newblock URL \url{https://www.nature.com/articles/s41467-021-21728-w}.

\bibitem[Sharma et~al.(2022)Sharma, Cerezo, Cincio, and
  Coles]{Sharma2022Trainability}
Kunal Sharma, M.~Cerezo, Lukasz Cincio, and Patrick~J. Coles.
\newblock Trainability of dissipative perceptron-based quantum neural networks.
\newblock \emph{Phys. Rev. Lett.}, 128:\penalty0 180505, May 2022.
\newblock \doi{10.1103/PhysRevLett.128.180505}.
\newblock URL \url{https://link.aps.org/doi/10.1103/PhysRevLett.128.180505}.

\bibitem[Patti et~al.(2021)Patti, Najafi, Gao, and
  Yelin]{patti_entanglement_2021}
Taylor~L. Patti, Khadijeh Najafi, Xun Gao, and Susanne~F. Yelin.
\newblock Entanglement devised barren plateau mitigation.
\newblock \emph{Phys. Rev. Res.}, 3\penalty0 (3):\penalty0 033090, July 2021.
\newblock ISSN 2643-1564.
\newblock \doi{10.1103/PhysRevResearch.3.033090}.
\newblock URL \url{https://link.aps.org/doi/10.1103/PhysRevResearch.3.033090}.

\bibitem[Bazant et~al.(1997)Bazant, Kaxiras, and Justo]{Bazant_Si}
Martin~Z. Bazant, Efthimios Kaxiras, and J.~F. Justo.
\newblock Environment-dependent interatomic potential for bulk silicon.
\newblock \emph{Phys. Rev. B}, 56:\penalty0 8542--8552, Oct 1997.
\newblock \doi{10.1103/PhysRevB.56.8542}.
\newblock URL \url{https://link.aps.org/doi/10.1103/PhysRevB.56.8542}.

\bibitem[Tersoff(1988)]{Tersoff_Si}
J.~Tersoff.
\newblock Empirical interatomic potential for silicon with improved elastic
  properties.
\newblock \emph{Phys. Rev. B}, 38:\penalty0 9902--9905, Nov 1988.
\newblock \doi{10.1103/PhysRevB.38.9902}.
\newblock URL \url{https://link.aps.org/doi/10.1103/PhysRevB.38.9902}.

\bibitem[O'Brien et~al.(2022)O'Brien, Streif, Rubin, Santagati, Su, Huggins,
  Goings, Moll, Kyoseva, Degroote, Tautermann, Lee, Berry, Wiebe, and
  Babbush]{obrien_efficient_2022}
Thomas~E. O'Brien, Michael Streif, Nicholas~C. Rubin, Raffaele Santagati, Yuan
  Su, William~J. Huggins, Joshua~J. Goings, Nikolaj Moll, Elica Kyoseva,
  Matthias Degroote, Christofer~S. Tautermann, Joonho Lee, Dominic~W. Berry,
  Nathan Wiebe, and Ryan Babbush.
\newblock Efficient quantum computation of molecular forces and other energy
  gradients.
\newblock \emph{Phys. Rev. Res.}, 4\penalty0 (4):\penalty0 043210, December
  2022.
\newblock \doi{10.1103/PhysRevResearch.4.043210}.
\newblock URL \url{https://link.aps.org/doi/10.1103/PhysRevResearch.4.043210}.

\bibitem[Lippert et~al.(1997)Lippert, Hutter, and Parrinello]{GPW1997}
GERALD Lippert, JURG Hutter, and MICHELE Parrinello.
\newblock A hybrid gaussian and plane wave density functional scheme.
\newblock \emph{Mol. Phys.}, 92\penalty0 (3):\penalty0 477--488, 1997.
\newblock \doi{10.1080/002689797170220}.

\bibitem[Goedecker et~al.(1996)Goedecker, Teter, and Hutter]{GTH1996}
S.~Goedecker, M.~Teter, and J.~Hutter.
\newblock Separable dual-space gaussian pseudopotentials.
\newblock \emph{Phys. Rev. B}, 54:\penalty0 1703--1710, Jul 1996.
\newblock \doi{10.1103/PhysRevB.54.1703}.
\newblock URL \url{https://link.aps.org/doi/10.1103/PhysRevB.54.1703}.

\bibitem[Hartwigsen et~al.(1998)Hartwigsen, Goedecker, and Hutter]{HGH1998}
C.~Hartwigsen, S.~Goedecker, and J.~Hutter.
\newblock Relativistic separable dual-space gaussian pseudopotentials from h to
  rn.
\newblock \emph{Phys. Rev. B}, 58:\penalty0 3641--3662, Aug 1998.
\newblock \doi{10.1103/PhysRevB.58.3641}.
\newblock URL \url{https://link.aps.org/doi/10.1103/PhysRevB.58.3641}.

\bibitem[Perdew et~al.(1996)Perdew, Burke, and Ernzerhof]{PBE1996}
John~P. Perdew, Kieron Burke, and Matthias Ernzerhof.
\newblock Generalized gradient approximation made simple.
\newblock \emph{Phys. Rev. Lett.}, 77:\penalty0 3865--3868, Oct 1996.
\newblock \doi{10.1103/PhysRevLett.77.3865}.
\newblock URL \url{https://link.aps.org/doi/10.1103/PhysRevLett.77.3865}.

\bibitem[VandeVondele et~al.(2005)VandeVondele, Krack, Mohamed, Parrinello,
  Chassaing, and Hutter]{VANDEVONDELE2005103}
Joost VandeVondele, Matthias Krack, Fawzi Mohamed, Michele Parrinello, Thomas
  Chassaing, and Jürg Hutter.
\newblock Quickstep: Fast and accurate density functional calculations using a
  mixed gaussian and plane waves approach.
\newblock \emph{Comput. Phys. Commun.}, 167\penalty0 (2):\penalty0 103--128,
  2005.
\newblock ISSN 0010-4655.
\newblock \doi{https://doi.org/10.1016/j.cpc.2004.12.014}.
\newblock URL
  \url{https://www.sciencedirect.com/science/article/pii/S0010465505000615}.

\bibitem[Kühne et~al.(2020)Kühne, Iannuzzi, Del~Ben, Rybkin, Seewald, Stein,
  Laino, Khaliullin, Schütt, Schiffmann, Golze, Wilhelm, Chulkov,
  Bani-Hashemian, Weber, Bor\v{s}tnik, Taillefumier, Jakobovits, Lazzaro,
  Pabst, Müller, Schade, Guidon, Andermatt, Holmberg, Schenter, Hehn, Bussy,
  Belleflamme, Tabacchi, Glöß, Lass, Bethune, Mundy, Plessl, Watkins,
  VandeVondele, Krack, and Hutter]{cp2k2020}
Thomas~D. Kühne, Marcella Iannuzzi, Mauro Del~Ben, Vladimir~V. Rybkin, Patrick
  Seewald, Frederick Stein, Teodoro Laino, Rustam~Z. Khaliullin, Ole Schütt,
  Florian Schiffmann, Dorothea Golze, Jan Wilhelm, Sergey Chulkov,
  Mohammad~Hossein Bani-Hashemian, Valéry Weber, Urban Bor\v{s}tnik, Mathieu
  Taillefumier, Alice~Shoshana Jakobovits, Alfio Lazzaro, Hans Pabst, Tiziano
  Müller, Robert Schade, Manuel Guidon, Samuel Andermatt, Nico Holmberg,
  Gregory~K. Schenter, Anna Hehn, Augustin Bussy, Fabian Belleflamme, Gloria
  Tabacchi, Andreas Glöß, Michael Lass, Iain Bethune, Christopher~J. Mundy,
  Christian Plessl, Matt Watkins, Joost VandeVondele, Matthias Krack, and Jürg
  Hutter.
\newblock {CP2K: An electronic structure and molecular dynamics software
  package - Quickstep: Efficient and accurate electronic structure
  calculations}.
\newblock \emph{J. Chem. Phys.}, 152\penalty0 (19):\penalty0 194103, 05 2020.
\newblock ISSN 0021-9606.
\newblock \doi{10.1063/5.0007045}.
\newblock URL \url{https://doi.org/10.1063/5.0007045}.

\bibitem[Swope et~al.(1982)Swope, Andersen, Berens, and
  Wilson]{velocityVerlet1982}
William~C. Swope, Hans~C. Andersen, Peter~H. Berens, and Kent~R. Wilson.
\newblock {A computer simulation method for the calculation of equilibrium
  constants for the formation of physical clusters of molecules: Application to
  small water clusters}.
\newblock \emph{J. Chem. Phys.}, 76\penalty0 (1):\penalty0 637--649, 01 1982.
\newblock ISSN 0021-9606.
\newblock \doi{10.1063/1.442716}.
\newblock URL \url{https://doi.org/10.1063/1.442716}.

\bibitem[Martyna et~al.(1992)Martyna, Klein, and Tuckerman]{NHC1992}
Glenn~J. Martyna, Michael~L. Klein, and Mark Tuckerman.
\newblock {Nos{\'e}--Hoover chains: The canonical ensemble via continuous
  dynamics}.
\newblock \emph{J. Chem. Phys.}, 97\penalty0 (4):\penalty0 2635--2643, 08 1992.
\newblock ISSN 0021-9606.
\newblock \doi{10.1063/1.463940}.
\newblock URL \url{https://doi.org/10.1063/1.463940}.

\bibitem[Geiger et~al.(2022)Geiger, Smidt, M., Miller, Boomsma, Dice,
  Lapchevskyi, Weiler, Tyszkiewicz, Batzner, Madisetti, Uhrin, Frellsen, Jung,
  Sanborn, Wen, Rackers, Rød, and Bailey]{e3nn}
Mario Geiger, Tess Smidt, Alby M., Benjamin~Kurt Miller, Wouter Boomsma,
  Bradley Dice, Kostiantyn Lapchevskyi, Maurice Weiler, Micha\l{} Tyszkiewicz,
  Simon Batzner, Dylan Madisetti, Martin Uhrin, Jes Frellsen, Nuri Jung, Sophia
  Sanborn, Mingjian Wen, Josh Rackers, Marcel Rød, and Michael Bailey.
\newblock Euclidean neural networks: e3nn, April 2022.
\newblock URL \url{https://doi.org/10.5281/zenodo.6459381}.

\bibitem[Reddi et~al.(2018)Reddi, Kale, and Kumar]{reddi_convergence_2018}
Sashank~J. Reddi, Satyen Kale, and Sanjiv Kumar.
\newblock On the {Convergence} of {Adam} and {Beyond}.
\newblock In \emph{Proceedings of the 6th {International} {Conference} on
  {Learning} {Representations}}, Vancouver, BC, Canada, February 2018.
  OpenReview.
\newblock URL \url{https://openreview.net/forum?id=ryQu7f-RZ}.
\newblock https://openreview.net/group?id=ICLR.cc/2018/Conference.

\bibitem[Kingma and Ba(2017)]{kingma_adam_2017}
Diederik~P. Kingma and Jimmy Ba.
\newblock Adam: {A} {Method} for {Stochastic} {Optimization}, January 2017.
\newblock URL \url{https://arxiv.org/abs/1412.6980}.

\bibitem[Larsen et~al.(2017)Larsen, Mortensen, Blomqvist, Castelli,
  Christensen, Du\l{}ak, Friis, Groves, Hammer, Hargus, Hermes, Jennings,
  Jensen, Kermode, Kitchin, Kolsbjerg, Kubal, Kaasbjerg, Lysgaard, Maronsson,
  Maxson, Olsen, Pastewka, Peterson, Rostgaard, Schiøtz, Schütt, Strange,
  Thygesen, Vegge, Vilhelmsen, Walter, Zeng, and Jacobsen]{larsen_atomic_2017}
Ask~Hjorth Larsen, Jens~Jørgen Mortensen, Jakob Blomqvist, Ivano~E. Castelli,
  Rune Christensen, Marcin Du\l{}ak, Jesper Friis, Michael~N. Groves, Bjørk
  Hammer, Cory Hargus, Eric~D. Hermes, Paul~C. Jennings, Peter~Bjerre Jensen,
  James Kermode, John~R. Kitchin, Esben~Leonhard Kolsbjerg, Joseph Kubal,
  Kristen Kaasbjerg, Steen Lysgaard, Jón~Bergmann Maronsson, Tristan Maxson,
  Thomas Olsen, Lars Pastewka, Andrew Peterson, Carsten Rostgaard, Jakob
  Schiøtz, Ole Schütt, Mikkel Strange, Kristian~S. Thygesen, Tejs Vegge,
  Lasse Vilhelmsen, Michael Walter, Zhenhua Zeng, and Karsten~W. Jacobsen.
\newblock The atomic simulation environment---a {Python} library for working
  with atoms.
\newblock \emph{J. Phys. Condens. Matter}, 29\penalty0 (27):\penalty0 273002,
  June 2017.
\newblock ISSN 0953-8984.
\newblock \doi{10.1088/1361-648X/aa680e}.
\newblock URL \url{https://dx.doi.org/10.1088/1361-648X/aa680e}.
\newblock Publisher: IOP Publishing.

\bibitem[Melchionna et~al.(1993)Melchionna, Ciccotti, and
  Holian]{melchionnaHooverNPTDynamics1993}
Simone Melchionna, Giovanni Ciccotti, and Brad~L. Holian.
\newblock Hoover {{NPT}} dynamics for systems varying in shape and size.
\newblock \emph{Mol. Phys.}, 78:\penalty0 533--544, 1993.

\bibitem[Melchionna(2000)]{melchionnaConstrainedSystemsStatistical2000}
Simone Melchionna.
\newblock Constrained systems and statistical distribution.
\newblock \emph{Phys. Rev. E}, 61\penalty0 (6):\penalty0 6165--6170, June 2000.
\newblock ISSN 1063-651X, 1095-3787.
\newblock \doi{10.1103/PhysRevE.61.6165}.

\bibitem[Holian et~al.(1990)Holian, De~Groot, Hoover, and
  Hoover]{holianTimereversibleEquilibriumNonequilibrium1990}
Brad~Lee Holian, Anthony~J. De~Groot, William~G. Hoover, and Carol~G. Hoover.
\newblock Time-reversible equilibrium and nonequilibrium isothermal-isobaric
  simulations with centered-difference {{Stoermer}} algorithms.
\newblock \emph{Phys. Rev. A}, 41\penalty0 (8):\penalty0 4552--4553, April
  1990.
\newblock ISSN 1050-2947, 1094-1622.
\newblock \doi{10.1103/PhysRevA.41.4552}.

\bibitem[Hastings(1970)]{MC}
W.~K. Hastings.
\newblock Monte carlo sampling methods using markov chains and their
  applications.
\newblock \emph{Biometrika}, 57\penalty0 (1):\penalty0 97--109, 04 1970.
\newblock ISSN 0006-3444.
\newblock \doi{10.1093/biomet/57.1.97}.
\newblock URL \url{https://doi.org/10.1093/biomet/57.1.97}.

\end{thebibliography}


\begin{thebibliography}{0}%
\makeatletter
\providecommand \@ifxundefined [1]{%
 \@ifx{#1\undefined}
}%
\providecommand \@ifnum [1]{%
 \ifnum #1\expandafter \@firstoftwo
 \else \expandafter \@secondoftwo
 \fi
}%
\providecommand \@ifx [1]{%
 \ifx #1\expandafter \@firstoftwo
 \else \expandafter \@secondoftwo
 \fi
}%
\providecommand \natexlab [1]{#1}%
\providecommand \enquote  [1]{``#1''}%
\providecommand \bibnamefont  [1]{#1}%
\providecommand \bibfnamefont [1]{#1}%
\providecommand \citenamefont [1]{#1}%
\providecommand \href@noop [0]{\@secondoftwo}%
\providecommand \href [0]{\begingroup \@sanitize@url \@href}%
\providecommand \@href[1]{\@@startlink{#1}\@@href}%
\providecommand \@@href[1]{\endgroup#1\@@endlink}%
\providecommand \@sanitize@url [0]{\catcode `\\12\catcode `\$12\catcode
  `\&12\catcode `\#12\catcode `\^12\catcode `\_12\catcode `\%12\relax}%
\providecommand \@@startlink[1]{}%
\providecommand \@@endlink[0]{}%
\providecommand \url  [0]{\begingroup\@sanitize@url \@url }%
\providecommand \@url [1]{\endgroup\@href {#1}{\urlprefix }}%
\providecommand \urlprefix  [0]{URL }%
\providecommand \Eprint [0]{\href }%
\providecommand \doibase [0]{https://doi.org/}%
\providecommand \selectlanguage [0]{\@gobble}%
\providecommand \bibinfo  [0]{\@secondoftwo}%
\providecommand \bibfield  [0]{\@secondoftwo}%
\providecommand \translation [1]{[#1]}%
\providecommand \BibitemOpen [0]{}%
\providecommand \bibitemStop [0]{}%
\providecommand \bibitemNoStop [0]{.\EOS\space}%
\providecommand \EOS [0]{\spacefactor3000\relax}%
\providecommand \BibitemShut  [1]{\csname bibitem#1\endcsname}%
\let\auto@bib@innerbib\@empty
\end{thebibliography}%

\end{document}


\title{Electronic Supplementary Information: Hybrid Quantum–Classical Machine Learning Potential with Variational Quantum Circuits}

\author{Soohaeng Yoo Willow}
\affiliation{Department of Energy Science, Sungkyunkwan University, Seobu-ro 2066, Suwon, 16419, Korea}

\author{David ChangMo Yang}
\affiliation{Department of Energy Science, Sungkyunkwan University, Seobu-ro 2066, Suwon, 16419, Korea}

\author{Chang Woo Myung}
\email{cwmyung@skku.edu}
\affiliation{Department of Energy Science, Sungkyunkwan University, Seobu-ro 2066, Suwon, 16419, Korea}
\affiliation{Department of Energy, Sungkyunkwan University, Seobu-ro 2066, Suwon, 16419, Korea}
\affiliation{Department of Quantum Information Engineering, Sungkyunkwan University, Seobu-ro 2066, Suwon, 16419, Korea}
\date{\today}

\maketitle

\newpage
\section{Rotationally equivariant tensor product}

The tensor product $\otimes$ in the interatomic continuous-filter convolution $f_\mathrm{Conv}$ is given by:
\begin{equation}
    [e_{ij}]_{m_e}^{l_e} \otimes [h_j]_{m_h}^{l_h} = \sum_{m_e, m_h} C_{l_h, m_h, l_e, m_e}^{l_o, m_o} [e_{ij}]_{m_e}^{l_e} [h_j]_{m_h}^{l_h} \rightarrow [O_{ij}]_{m_o}^{l_o},
\end{equation}
where ``rotational order'' $l = 0, 1, 2, \dots$ is a non-negative integer.
The representation index $m$ takes values $m \in [-l, l]$.
$C$ indicates the Clebsch-Gordan coefficients. 

\newpage 

\begin{figure} [htp]
\centering
\includegraphics[width=\linewidth]{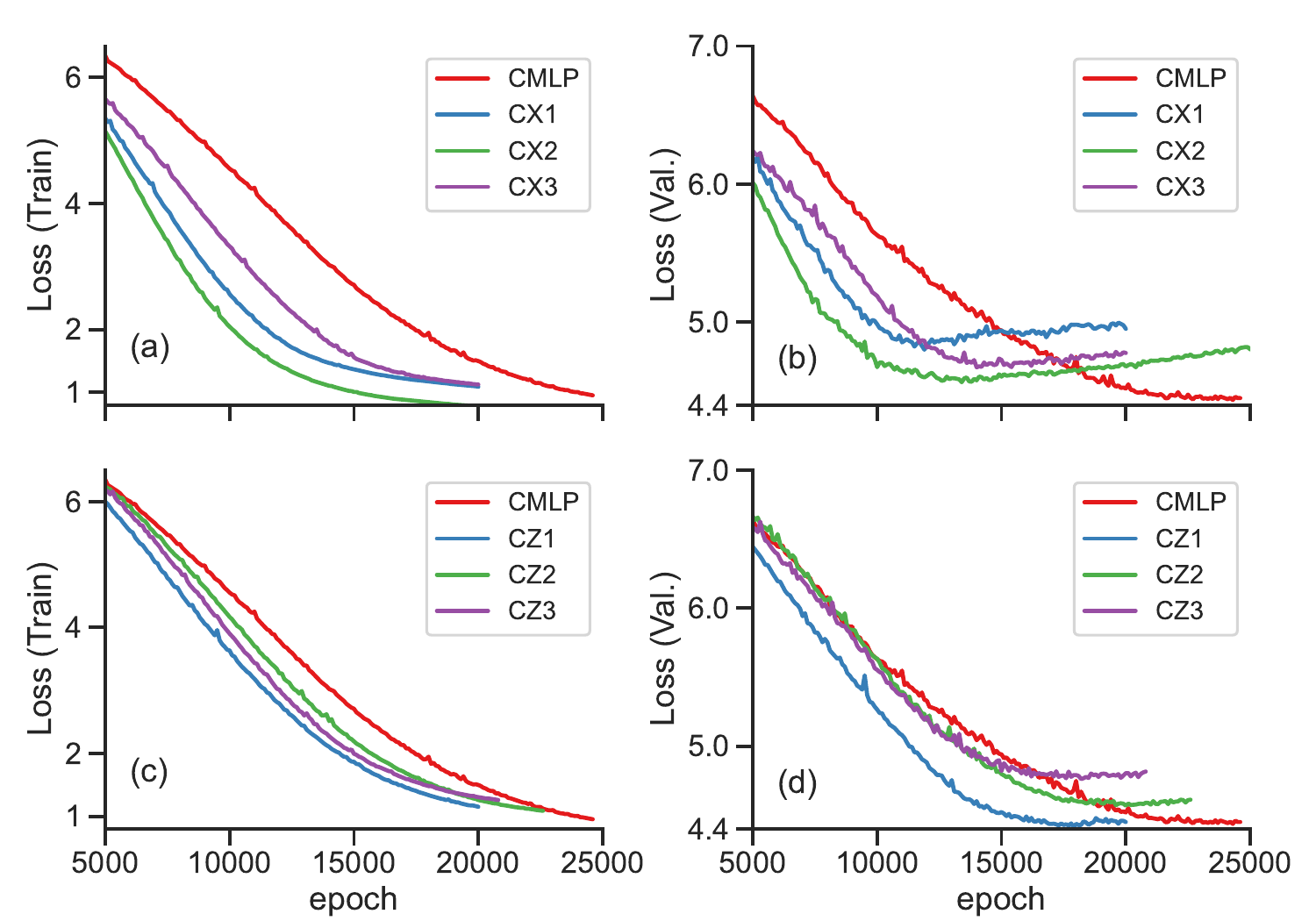}
\caption{The loss profiles of hybrid quantum-classical machine learning potential (HQC-MLP) models with \texttt{BasicEntanglerCX} and \texttt{BasicEntaglerCZ} Ansatz modules compared to a classical machine learning potential (CMLP). Panels (a) and (b) show training and validation losses, respectively, for the \texttt{BasicEntanglerCX} Ansatz with three different random seeds (CX1, CX2, CX3). Panels (c) and (d) show corresponding results for the \texttt{BasicEntanglerCZ} Ansatz (CZ1, CZ2, CZ3). In the validation datasets, the loss values exhibit clear minima, indicating the onset of overfitting as training progresses.}

\end{figure}

\newpage
\begin{figure} [htp]
\centering
\includegraphics[width=\linewidth]{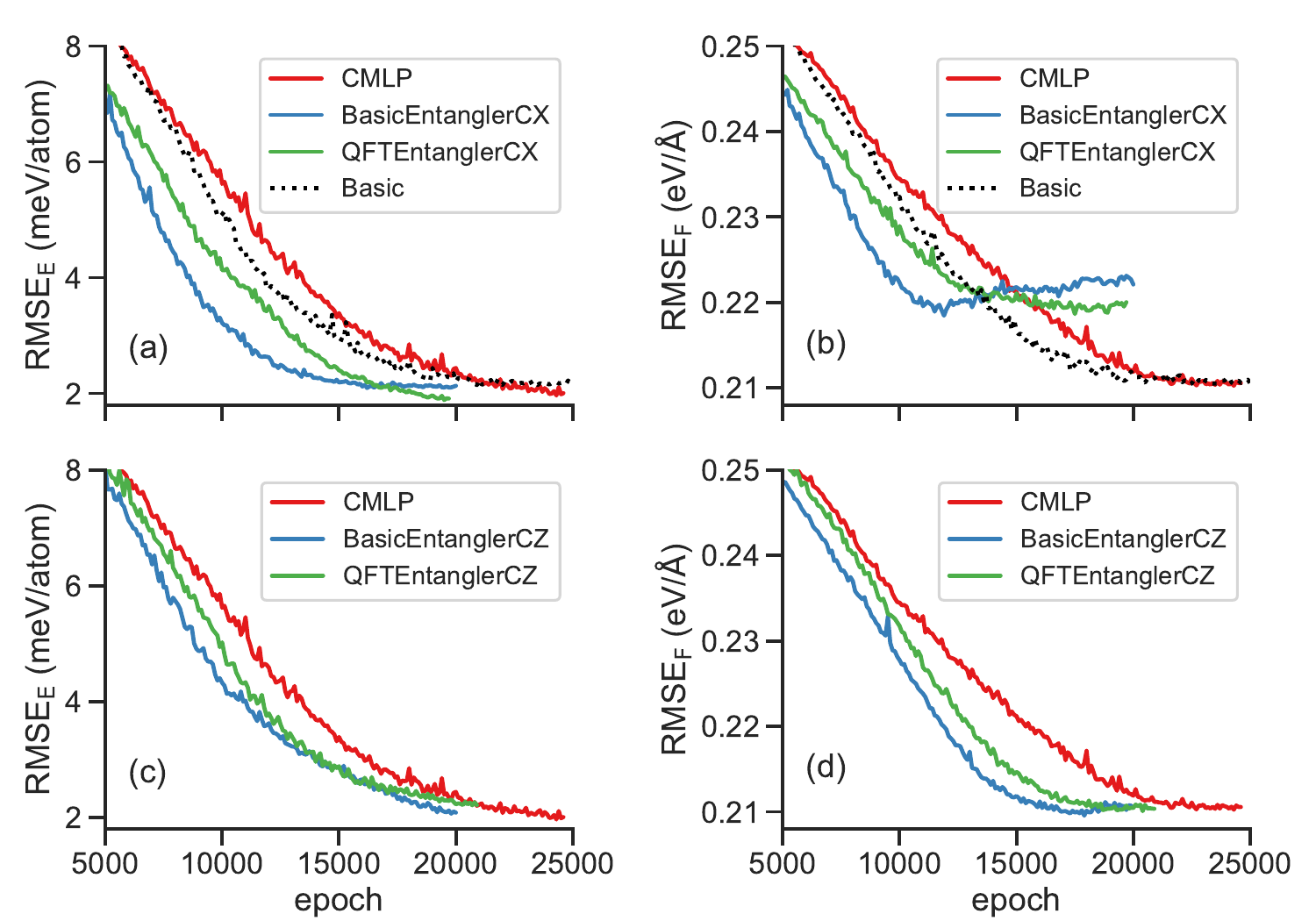}
\caption{The root mean squared errors (RMSEs) for predicted energies per atom (RMSE$_\mathrm{E}$) and predicted forces (RMSE$_\mathrm{F}$) are shown using the test dataset. 
The HQC-MLP model (blue bashed lines)  shows a faster decrease in both RMSE$_\mathrm{E}$ and RMSE$_\mathrm{F}$ compared to the CMLP model (red lines). 
Despite some overfitting in HQC-MLP model, 
both models produced similar values: RMSE$_\mathrm{E}$ $\approx$ 2.0 meV/atom and RMSE$_\mathrm{F}$ $\approx$ 0.21 $\sim$ 0.22 eV/\AA.}

\end{figure}

\newpage
\begin{figure} [htp]
\centering
\includegraphics[width=0.8\linewidth]{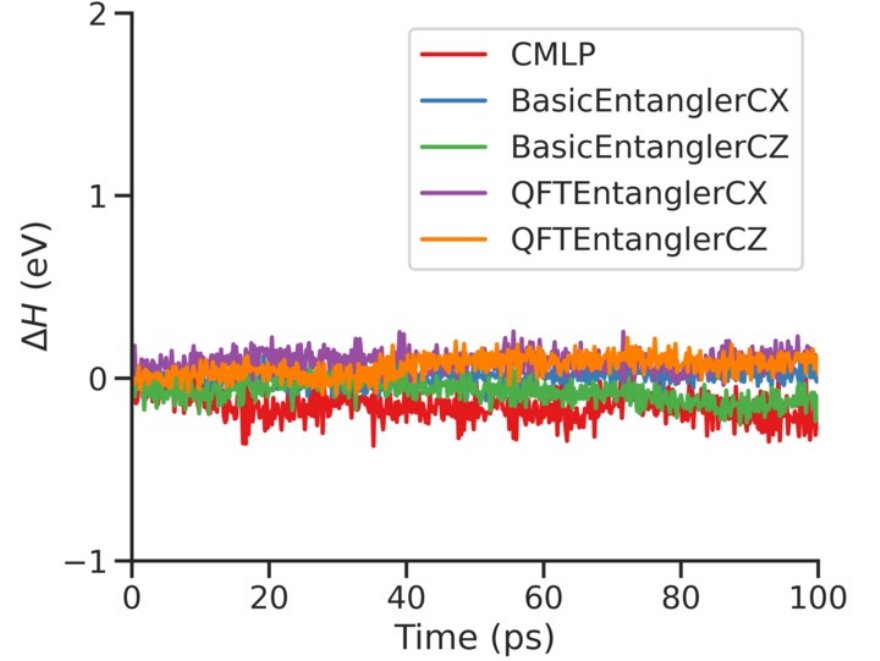}
\caption{Conservation of the Hamiltonian. Canonical molecular dynamics simulations were carried out at a temperature of $T=2000$ K using machine learning potential models. 
The relative Hamiltonian, denoted as $\Delta H$, is defined by the expression $\Delta H = H(t) - H(t=0)$, where $H(t)$ represents the Hamiltonian at time $t$ and $H(0)$ is the initial Hamiltonian.}

\end{figure}

\newpage
\begin{figure} [htp]
\centering
\includegraphics[width=0.8\linewidth]{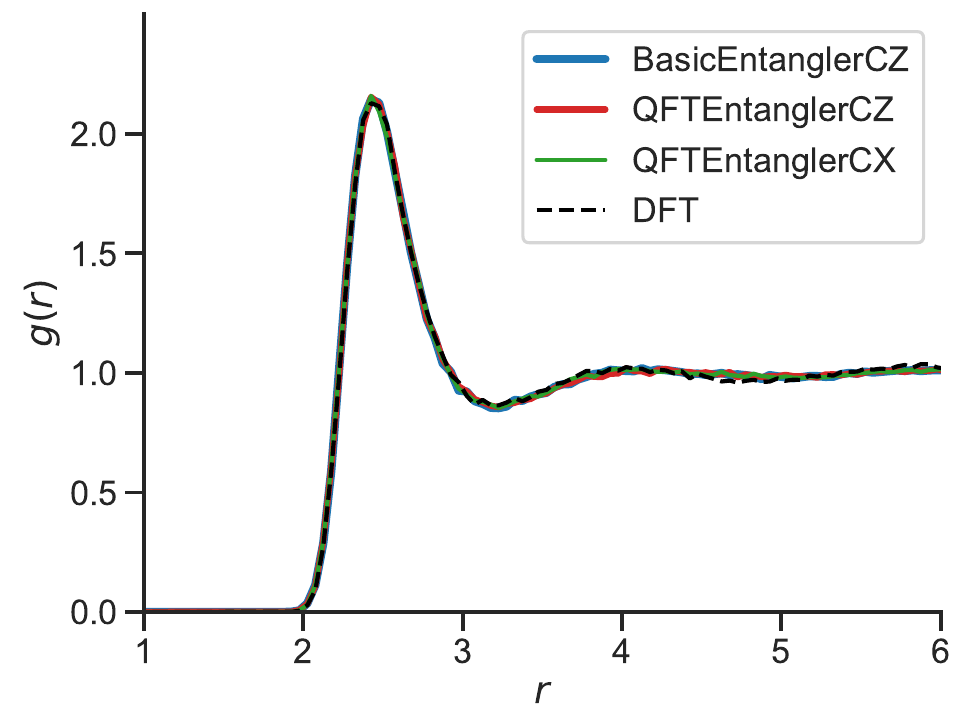}
\caption{Radial distribution function, $g(r)$, for liquid silicon at $T = 2000$ K. The plot compares the results from the HQC-MLP (\texttt{BasicEntanglerCZ}, \texttt{QFTEntanglerCZ}, and \texttt{QFTEntanglerCX}) with the reference calculated using DFT (dashed black line), demonstrating excellent agreement.}

\end{figure}

\label{fig:H}